\documentclass[lettersize,journal]{IEEEtran}
\usepackage{amssymb}

\ifCLASSINFOpdf
\else
\fi

\usepackage{stfloats}
\usepackage{amsmath,amsfonts}
\usepackage{algorithmic}
\usepackage{algorithm}
\usepackage{array}
\usepackage[caption=false,font=normalsize,labelfont=sf,textfont=sf]{subfig}
\usepackage{textcomp}
\usepackage{mathtools}
\usepackage{stfloats}
\usepackage{url}
\usepackage{verbatim}
\usepackage{graphicx}
\usepackage{cite}
\usepackage{times}
\usepackage{epsfig}
\usepackage{multirow}
\usepackage{multicol}
\usepackage{color}
\usepackage{yhmath}
\usepackage{bm}
\usepackage{pifont}
\usepackage{eso-pic}
\usepackage{epsfig}
\usepackage{bbding}
\usepackage[autostyle=true]{csquotes}
\usepackage{booktabs}
\usepackage{url}
\usepackage[
    colorlinks=true,
    linkcolor=red, 
    citecolor=green
]{hyperref}
\usepackage{tikz,xcolor} 
\usepackage{cite}
\usepackage{amsmath}
\usepackage{bbding}
\usepackage{cleveref} 
\usepackage[table]{xcolor}
\usepackage{pifont}
\usepackage{multirow}
\definecolor{lime}{HTML}{A6CE39}
\usepackage{graphicx} 
\DeclareRobustCommand{\orcidicon}{%
    \begin{tikzpicture}
    \draw[lime, fill=lime] (0,0) 
    circle [radius=0.16] 
    node[white] {{\fontfamily{qag}\selectfont \small ID}};    \draw[white, fill=white] (-0.0625,0.095) 
    circle [radius=0.007];    \end{tikzpicture}
    \hspace{-2mm}}
\foreach \x in {A, ..., Z}{%
    \expandafter\xdef\csname orcid\x\endcsname{\noexpand\href{https://orcid.org/\csname orcidauthor\x\endcsname}{\noexpand\orcidicon}}
}

\definecolor{mycolor}{RGB}{0, 100, 200} 
\colorlet{mycolor}{black}
\newcommand{\myrevision}[1]{\textcolor{mycolor}{#1}}

\newcommand{\myright}{\textcolor{black}{\ding{51}}}
\newcommand{\mywrong}{\textcolor{black}{\ding{55}}}

\newlength{\captiongap}
\begin{document}
%
\title{Symmetric Entropy-Constrained\\Video Coding for Machines}

\author{
Yuxiao Sun\orcidA{}, 
Meiqin Liu\orcidC{}, 
Chao Yao\orcidD{},~\IEEEmembership{Member,~IEEE},
Qi Tang\orcidH{}, 
Jian Jin\orcidE{},~\IEEEmembership{Member,~IEEE}, \\
Weisi Lin\orcidG{},~\IEEEmembership{Fellow,~IEEE},
Frederic Dufaux\orcidX{},~\IEEEmembership{Fellow,~IEEE},
and Yao Zhao\orcidB{},~\IEEEmembership{Fellow,~IEEE}
\thanks{
This work was supported in part by the National Natural Science Foundation of China under Grant 62120106009, Grant 62372036, Grant U24B20179, and Grant 62332017. \textit{(Yuxiao Sun and Meiqin Liu contributed equally to this work.)} \textit{(Corresponding author: Yao Zhao.)} 

Yuxiao Sun, Meiqin Liu, Qi Tang, and Yao Zhao are with the Institute of Information Science, Beijing Jiaotong University, Beijing 100044, China, and also with the Visual Intelligence +X International Cooperation Joint Laboratory of MOE, Beijing 100044, China (e-mail: \url{yuxiaosun@bjtu.edu.cn}; \url{mqliu@bjtu.edu.cn}; \url{qitang@bjtu.edu.cn}; \url{yzhao@bjtu.edu.cn}).

Chao Yao is with the School of Computer and Communication Engineering, University of Science and Technology Beijing, Beijing 100083, China (e-mail: \url{yaochao@ustb.edu.cn}).

Jian Jin and Weisi Lin are with the School of Computer Science and Engineering, Nanyang Technological University, Singapore 639798, and also with the China-Singapore International Joint Research Institute (CSIJRI), Guangzhou 510555, China (e-mail: \url{jian.jin@ntu.edu.sg}; \url{wslin@ntu.edu.sg}).

Frederic Dufaux is with Laboratoire des Signaux et Systèmes, Université Paris-Saclay, CNRS, CentraleSupelec, 91192, Gif-sur-Yvette, France (e-mail: \url{frederic.dufaux@l2s.centralesupelec.fr}). 

}
}
%
%

\markboth{This is the author's version (AAM version) accepted for publication in IEEE Transactions on Image Processing.}%
{This is the author's version (AAM version) accepted for publication in IEEE Transactions on Image Processing.}
%


\IEEEoverridecommandlockouts
\makeatletter
\def\ps@IEEEtitlepagestyle{%
  \def\@oddhead{\mbox{}\hfill\thepage}%
  \def\@evenhead{\thepage\hfill\mbox{}}%
  \def\@oddfoot{\parbox{\textwidth}{\footnotesize\copyright~2026 IEEE. Personal use of this material is permitted. Permission from IEEE must be obtained for all other uses, in any current or future media, including reprinting/republishing this material for advertising or promotional purposes, creating new collective works, for resale or redistribution to servers or lists, or reuse of any copyrighted component of this work in other works. DOI: 10.1109/TIP.2026.3705185}}%
  \def\@evenfoot{}%
}
\makeatother

\maketitle
\thispagestyle{IEEEtitlepagestyle}

\begin{abstract}

As video transmission increasingly serves machine vision systems (MVS) instead of human vision systems (HVS), video coding for machines (VCM) has become a critical research topic. Existing VCM methods often bind codecs to specific downstream models, requiring retraining or supervised data, thus limiting generalization in multi-task scenarios. Recently, unified VCM frameworks have employed visual backbones (VB) and visual foundation models (VFM) to support multiple video understanding tasks with a single codec. 
They mainly utilize VB/VFM to maintain semantic consistency or suppress non-semantic information, but seldom explore how to directly link video coding with understanding under VB/VFM guidance.
Hence, we propose a Symmetric Entropy-Constrained Video Coding framework for Machines (SEC-VCM). 
It establishes a symmetric alignment between the video codec and VB, allowing the codec to leverage VB's representation capabilities to preserve semantics and discard MVS-irrelevant information.
Specifically, a bi-directional entropy-constraint (BiEC) mechanism ensures symmetry between the process of video decoding and VB encoding by suppressing conditional entropy. 
This helps the codec to explicitly handle semantic information beneficial to MVS while squeezing useless information. 
Furthermore, a semantic-pixel dual-path fusion (SPDF) module injects pixel-level priors into the final reconstruction. 
Through semantic-pixel fusion, it suppresses artifacts harmful to MVS and improves machine-oriented reconstruction quality. 
\myrevision{Experimental results on classical video understanding tasks and MLLM-based tasks show state-of-the-art~(SOTA) rate-task performance. It achieves significant bitrate savings over H.266/VVC reference software VTM on video instance segmentation~(37.4\%), video object segmentation~(29.8\%), object detection~(46.2\%), multiple object tracking~(44.9\%), and MLLM-based video grounding~(97.6\%).} Our code is in \url{https://github.com/Ws-Syx/SEC-VCM}. 
\end{abstract}


\section{Introduction}

\begin{figure}[!t]
\centering
\includegraphics[width=\linewidth]{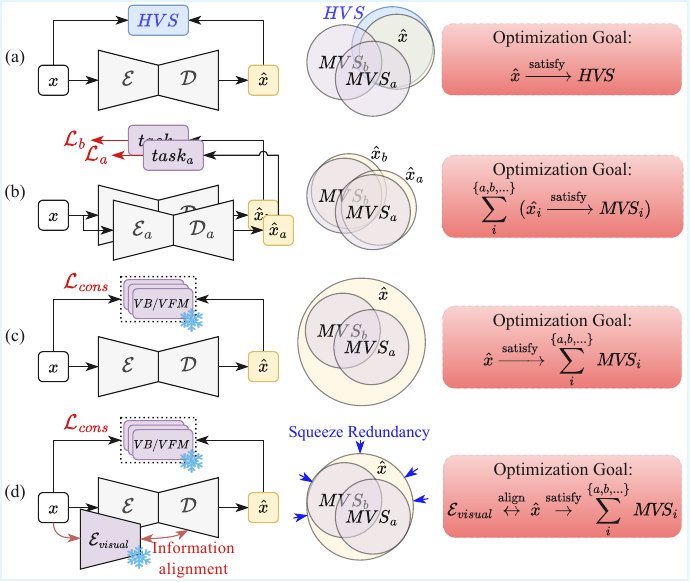}
\caption{\myrevision{Overview of different methods. $\mathcal{L}_a$ and $\mathcal{L}_b$ denote downstream task losses, $\mathcal{L}_{\textit{cons}}$ denotes loss for maintaining feature consistency, and $\mathcal{E}_{\textit{visual}}$ denotes the encoder of visual backbone (VB) or visual foundation model (VFM). (a) Standard or neural video codecs~\cite{h264, h265} are optimized for HVS, which makes the decoded videos $\hat{x}$ suboptimal for machine vision. (b) Each neural video/feature codec~\cite{sfma, sheng2023lvvc, misra2022video} is optimized for a specific machine vision task by downstream task loss. (c) The unified codec~\cite{smc, smc++, sa_icm, free_vsc} is optimized with the assistance of the VB or VFM to reconstruct videos suitable for multiple downstream tasks. (d) Our unified compression framework aligns the video coding process with the video understanding process from the entropy perspective and squeezes machine-vision-irrelevant information during the process of video reconstruction.} }
\label{figure-first}
\end{figure}

\IEEEPARstart{D}{igital} video plays a crucial role in our daily lives, comprising a significant portion of data storage. \myrevision{The growing prevalence of automated video analytics (e.g., surveillance, autonomous driving, and content moderation) has shifted the consumer of compressed videos from humans to machines, yet mainstream codecs are still optimized for the human visual system and thus spend bits on details that machines often ignore.} To efficiently store and transmit this massive amount of data, video coding techniques have been developed, primarily designed to meet the requirements of the human visual system (HVS). For videos aimed at HVS, such as movies and short clips, it is essential to preserve visual details perceptible to humans during compression. Meanwhile, videos collected for machine vision tasks, such as surveillance analysis~\cite{PANDEESWARI2025111349} and facial recognition~\cite{facial}, typically require only part of the video information, rather than the complete visual details~\cite{scalable_choi, scalable_dcc}. Currently, neural video coding frameworks have evolved significantly and offer excellent compression performance~\cite{dcvc_rt, dcvc_fm, dcvc_dc, dcvc_sdd, ibvc, ecvc}. However, these methods are optimized primarily for pixel-domain distortion or HVS-related metrics and do not fully satisfy the requirements of machine vision systems (MVS) in terms of video understanding (e.g., detection, segmentation, and object tracking)~\cite{trans_vfc}, as shown in Fig.~\ref{figure-first}(a). Therefore, it is essential to develop video coding for machines (VCM) methods and bridge the gap between video coding and video understanding.

\myrevision{A natural first attempt is task-coupled supervised training.} To reconstruct machine-vision-friendly videos, most existing methods~\cite{end2endFeatureCompression_multiscale1, end2endFeatureCompression_multiscale2, sheng2023lvvc} bridge the upstream codec with specific downstream task models and jointly optimize with the loss functions of different tasks, as shown in Fig.~\ref{figure-first}(b). However, this requires retraining the entire upstream codec or downstream models for each new task, resulting in significant computational and time costs. Meanwhile, some parameter-efficient fine-tuning (PEFT) strategies are introduced to adapt codecs to new downstream tasks with lower cost~\cite{sfma}. Nonetheless, most of them still rely on the labeled data and supervised training process for specific downstream tasks, limiting their generalization in multi-task scenarios. 

\myrevision{A more scalable alternative is to let a single codec serve many downstream tasks without task-specific labels.} To improve the applicability, some methods~\cite{cta_ged, free_vsc} have adopted unified compression frameworks to reconstruct videos suitable for multiple downstream tasks simultaneously. These methods abandon supervised training and instead maintain perceptual consistency in a self-supervised manner. Since the visual backbone (VB) and visual foundation model (VFM), such as ResNet50~\cite{resnet50}, DINOv2~\cite{dinov2}, and Mask Auto-Encoder~(MAE)~\cite{video_mae}, are capable of extracting general feature representations for video understanding~\cite{cta_ged}, they are introduced to maintain semantic consistency between the original and reconstructed frames~\cite{smc, smc++, sa_icm, yin2025unified_clip}, as shown in Fig.~\ref{figure-first}(c). Additionally, some methods~\cite{smc, smc++} have extended an analysis-oriented bitstream beyond standard codecs (e.g., H.264/AVC~\cite{h264} and H.266/VVC~\cite{vvenc}) and introduced VB and VFM to suppress non-semantic information in the analysis-oriented bitstream. 
However, these approaches still treat VB/VFM merely as auxiliary tools for enforcing semantic consistency or non-semantic suppressing, without fully exploiting their potential to guide the video codec itself. This limitation leaves the video coding process loosely connected to the video understanding process by machines, motivating us to design a framework that explicitly aligns the two.

\myrevision{Our key insight is simple: if a pretrained visual backbone already defines what information machines care about, the codec should be symmetrically aligned with that backbone from an information-theoretic perspective, so that task-irrelevant information is squeezed.} To address the above issues, we propose a Symmetric Entropy-Constrained Video Coding framework for Machine~(SEC-VCM). It supports multiple downstream tasks with a single unsupervised training process and further eliminates machine-vision-irrelevant information by leveraging existing visual backbones, as shown in Fig.~\ref{figure-first}(d). Its key principle is to bridge the gap between the video coding and understanding processes from an entropy perspective, explicitly guiding the video codec to handle only information beneficial to video understanding while discarding irrelevant information. 
Specifically, an information-alignment-based semantic decoder with a bi-directional entropy-constraint (BiEC) mechanism is proposed to make the video decoding process of the codec symmetric and consistent with the encoding process of the visual backbone. 
In addition, a semantic-pixel dual-path fusion (SPDF) module is designed to incorporate HVS-related priors into the machine-oriented reconstruction process. It helps reduce artifacts harmful to MVS and fulfill the frame content through a gated fusion mechanism. 
The experiments on multiple machine vision tasks demonstrate that the proposed SEC-VCM achieves state-of-the-art (SOTA) rate-task performance. Compared to VTM, our method achieves significant bitrate reductions, with savings of 37.4\% for video instance segmentation, 29.8\% for video object segmentation, 46.2\% for object detection, 44.9\% for multiple object tracking, \myrevision{and 97.6\% for MLLM-based video grounding}.

Our main contributions are summarized as follows: 

\begin{itemize} 
    \item We propose SEC-VCM, a novel video coding framework explicitly aligned with machine vision requirements from an information-theoretic perspective, enabling general support for diverse downstream machine tasks.
    \item We introduce an information-alignment-based semantic decoder with a bi-directional entropy-constraint mechanism, which aligns the codec’s decoding process with the visual backbone’s encoding process, thereby eliminating machine-vision-irrelevant redundancy. 
    \item We design a semantic-pixel dual-path fusion module to inject pixel-level prior knowledge into the machine-oriented decoding process for artifact suppression, details fulfillment, and improved downstream tasks accuracy. 
    
\end{itemize}

\section{Related Works}

\subsection{Neural Video Coding}

Most existing neural video coding methods~\cite{dvc, dcvc, dcvc_hem, dcvc_dc, ssf, dcvc_sdd, ibvc, tmm2023deformable} adopt a two-step pipeline including inter-prediction and residual or context compression. 
Lu~\emph{et al.}~\cite{dvc} proposed the first neural video codec DVC, which uses the optical flow~\cite{spynet} for inter-prediction. 
Hu~\emph{et al.}~\cite{fvc} proposed to convert the processes of motion estimation (ME), motion compensation (MC), and residual coding from the pixel domain to the feature domain. 
Li~\emph{et al.}~\cite{dcvc} improved the concept of residual coding and proposed the conditional coding paradigm for better compression performance, termed DCVC. 
Sheng~\emph{et al.}~\cite{dcvc_tcm} upgraded the DCVC and proposed DCVC-TCM with a multi-scale conditional coding to better remove temporal redundancy.
Furthermore, many methods~\cite{dcvc_sdd, dcvc_fm, ecvc} introduce long-range training to suppress accumulation error and make the codec perform better on a long group of pictures (GOP). Overall, existing neural video coding methods improve compression efficiency from various perspectives, such as inter- or bi-directional prediction~\cite{ssf, dcvc_sdd, dcvc_dc, ibvc, plvc}, residual/context compression~\cite{dcvc_rt}, and entropy models~\cite{dcvc_hem, dcvc_rt}. Many NVC methods, such as ECVC-series~\cite{ecvc} and DCVC-series~\cite{dcvc_dc, dcvc_fm, dcvc_rt}, demonstrate excellent compression performance.

\subsection{\myrevision{Standard Video Coding for Machines}}

\myrevision{Besides the recent surge of end-to-end neural codecs, there are many standards-based VCM solutions~\cite{vvc_m, hevc_detection, cta_ged}. Early Moving Picture Experts Group (MPEG) standards, such as compact descriptors for visual search (CDVS)~\cite{cdvs} and compact descriptors for video analysis (CDVA)~\cite{cdva}, aimed to extract and compress hand-crafted or learned visual descriptors instead of preserving full pixel fidelity, thereby reducing transmission cost for visual analysis tasks. Additionally, recent studies have investigated the direct adaptation of standard codecs for VCM. For example, task-adaptive pre-processing modules can be inserted before standard codecs to modify input content toward downstream machine tasks, enabling one codec to better serve multiple downstream analyses~\cite{cta_preprocess3}. In parallel, just-recognizable-difference~(JRD)~\cite{liu2026_DT_JRD} extends the just-noticeable-difference~(JND) concept from human vision to machine vision systems, providing a deep transformer-based prediction model that enables VVC parameters determination. Meanwhile, satisfied-machine-ratio~(SMR)~\cite{cmjia_SMR} introduces a novel perceptual quality metric that aggregates satisfaction scores from diverse machine models, addressing the limitation of single-machine evaluation in previous works and promoting generality. More recently, MPEG has advanced VCM standardization from both feature compression and machine-friendly video compression perspectives, incorporating tools such as spatial/temporal resampling, region-of-interest~(ROI) processing, and bit-depth truncation on top of standard codecs~\cite{mpeg_vcm_roi, mpeg_vcm_roi2}. Nevertheless, these standard-codec-based approaches still rely heavily on hand-crafted tools, additional modules, and external optimization strategies, making it difficult to fully align the coding process with general video understanding. }

\subsection{Neural Video Coding for Machines}

The field of video coding for machines mainly includes two paradigms: analyze-then-compress (ATC) and compress-then-analyze (CTA). \myrevision{In addition, VCM can also be viewed as a sub-task of video coding for machines and humans (VCMH).}

\subsubsection{Analyze-then-Compress Paradigm}

Since many edge devices (e.g., Smartphones, AI cameras) support neural processing, they can extract features from images or videos, compress them, and transmit these features to reduce the computational load on the server. Most methods rely on labeled data and joint optimization strategies to make neural feature codecs meet the needs of specific downstream tasks~\cite {feature_compression_gcs, end2endFeatureCompression2, end2endFeatureCompression_multiscale1, end2endFeatureCompression_multiscale2, misra2022video}. However, these methods rely on retraining the upstream codec or downstream task models to adapt to newly coming downstream tasks. Some works introduced lightweight adapters for latent-domain~\cite{zhang2025perception} or feature-domain~\cite{trans_vfc} transformation, enabling the reconstructed features to satisfy multiple downstream tasks and avoiding heavy fine-tuning.

\begin{figure*}[t]
\centering
\includegraphics[width=\textwidth]{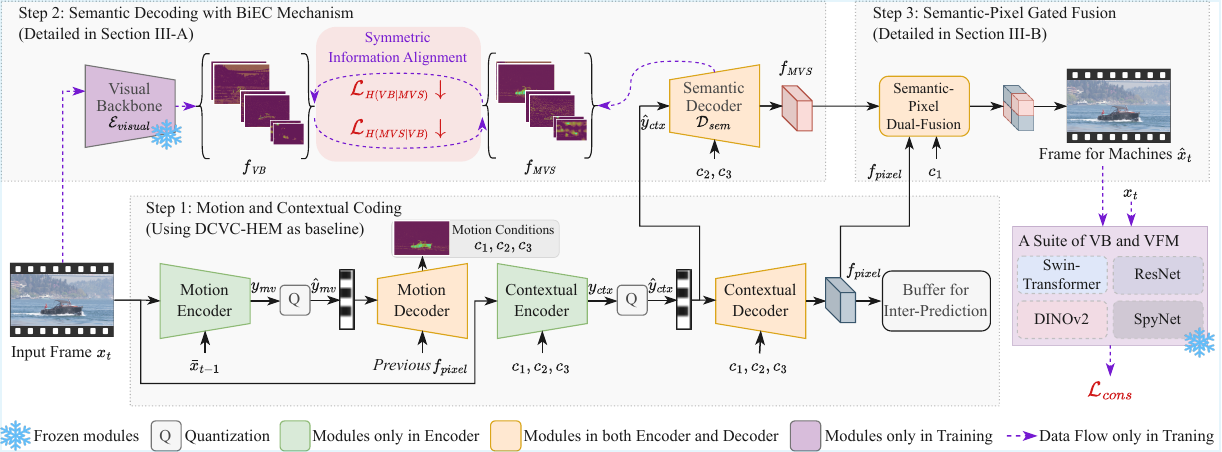}
\caption{\myrevision{Overview of the proposed SEC-VCM framework, \myrevision{which contains 3 main steps.} (1)~The basic compression loop contains motion and contextual coding processes. ``Q'' denotes the quantization operation. Notably, some modules in this step are not shown above for simplicity, since this step is directly adopted from the neural video codec DCVC-HEM~\cite{dcvc_hem}. (2)~The semantic decoding process is information-aligned with the encoding process of the pretrained visual backbone by the proposed BiEC mechanism from an entropy perspective. This helps the semantic decoder $\mathcal{D}_{\textit{sem}}$ purely handle information beneficial to MVS. (3)~The pixel-level prior knowledge $f_{\textit{pixel}}$ is injected into the perception-oriented decoding process to fulfill frame content, suppress artifacts, and enhance final reconstruction quality.} }
\label{figure-overview}
\end{figure*}

\subsubsection{Compress-then-Analyze Paradigm}

Some CTA methods cascade neural codecs or pre-process modules with downstream tasks and adopt jointly supervised training to achieve promising rate-task performance on specific tasks (e.g., object detection, face recognition, optical character recognition)~\cite{cta_joint, cta_preprocess, cta_preprocess2, cta_preprocess3}. To reduce training costs, some methods explored parameter-efficient fine-tuning (PEFT) based strategies of the upstream codecs for specific downstream tasks~\cite{sfma, trans_tic}. However, these methods need to fine-tune the entire codec or specific modules for each downstream task, which increases costs and limits their generalization. Meanwhile, some methods abandoned task-specific joint training and proposed unified VCM codecs generalizable to multiple downstream tasks. Tian~\emph{et al.} proposed to enhance edge~\cite{cta_ged} and semantic information~\cite{smc, smc++} by introducing an additional bitstream beyond the standard codecs. Other methods incorporated the visual backbone (VB) or the visual foundation model (VFM) for self-supervised training. Shindo~\emph{et al.} proposed the SA-ICM~\cite{sa_icm}, which uses the Segment Anything Model (SAM) to constrain the semantic and edge consistency between the original and reconstructed images. Tian~\emph{et al.}~\cite{free_vsc} employed DINOv2~\cite{dinov2} and VideoMAE~\cite{video_mae} in the latent domain to maintain semantic integrity. These methods typically apply VB and VFM to compute feature-domain losses for semantic consistency, yet they have not fully explored the alignment of video coding and video understanding at the information level, nor investigated how video understanding can explicitly guide the video coding process.
\subsubsection{\myrevision{Coding for Machines and Humans}}

\myrevision{A related topic to VCM is video/image coding for machines and humans~(VCMH/ICMH), which attempts to support human viewing and machine analysis simultaneously. Sheng~\emph{et al.} proposed the VNVC~\cite{sheng2023lvvc} framework, which optimizes a neural video codec for humans at the first training stage and freezes the codec and transforms intermediate features for multiple downstream tasks by fine-tuning downstream networks. Huang~\emph{et al.} proposed HMFVC~\cite{cmjia_huang2025_hmfvc}, which learns semantic representations shared by signal reconstruction and visual analysis, enabling joint optimization for compression, perception, and machine task accuracy. More recently, Feng~\emph{et al.} proposed Diff-ICMH~\cite{feng2025difficmh}, which leverages a diffusion model in the image decoding process to improve perceptual realism for HVS while maintaining semantic fidelity for machines. Most of VCM-related studies~\cite{scalable_choi, scalable_dcc, scalable_liu, vvc_m, yin2025unified_clip, deepsvc} suggest that HVS and MVS share some commonalities, and consider VCM as a sub-task of VCMH. In contrast, our work concentrates on a pure VCM setting and aims to improve machine-oriented video reconstruction for multiple downstream tasks without task-specific adaptation.}

\section{Methodology}

\subsection{Motivation and Overview}

\myrevision{Before the formal derivation, we summarize our framework in one sentence: \emph{We train a general machine-oriented video codec whose intermediate features reside in the same information-theoretic manifold as a pretrained visual backbone, and employs semantic-pixel fusion to restore low-level details that semantic information alone cannot reconstruct}. The rest of this section formalizes this idea.}

Existing neural video codecs typically adopt VAE-like networks~\cite{balle2016end} to compress the input video $x$ into compact latent representations and reconstruct the frame $\hat{x}$ for the machine vision system (MVS). Current methods optimize the codec for specific or multiple machine vision tasks, either in a supervised or unsupervised manner~\cite {sheng2023lvvc, misra2022video, smc, smc++}. Also, some methods maintain semantic consistency by minimizing the perceptual distance between the original video $x$ and the compressed video $\hat{x}$ measured by visual backbone (VB) or visual foundation model (VFM)~\cite{free_vsc, sa_icm}. Their paradigm is formulated as follows: 
\begin{equation}
\left\{
\begin{aligned}
    & \hat{x}=\mathcal{D}(\lfloor\mathcal{E}(x)\rceil), \\
    & \mathcal{L}=R + D(x, \hat{x}) + D(\mathcal{E}_{\textit{visual}}(x),\mathcal{E}_{\textit{visual}}(\hat{x})) + \mathcal{L}_{task(\hat{x})},
\end{aligned}
\right.
\end{equation}
\noindent where $\mathcal{E}(\cdot)$ and $\mathcal{D}(\cdot)$ denote the encoder and decoder of the video codec, ``$\lfloor \cdot \rceil$'' denotes the quantization operation~\cite{balle2016end}. $\mathcal{L}$ denotes the total loss, $R$ denotes the bitrate, $\mathcal{E}_{\textit{visual}}$ denotes the encoder of VB or VFM, $D(\cdot, \cdot)$ denotes pixel-domain or feature-domain distortion (e.g., MSE), and $\mathcal{L}_{task(\hat{x})}$ denotes the loss function of the downstream task.  




Although VB and VFM have been introduced to help codecs preserve semantics, their integration with video coding remains unexplored. They excel at extracting rich and general feature representations for multiple machine vision tasks~\cite{resnet50, dinov2}, yet such features are not naturally suited for entropy coding or direct compression. To better leverage their general capacity, we propose to align the video codec with VB/VFM from an entropy perspective, encouraging it to selectively preserve features beneficial to MVS while squeezing irrelevant information. The alignment objective is formulated as:

\begin{equation}
\mathcal{D}_{\textit{sem}}(\hat{y}) \xleftrightarrow{\text{aligned}} \mathcal{E}_{\textit{visual}}(x) \xrightarrow{\text{capable}} \text{downstream tasks},
\end{equation}

\noindent where $\mathcal{D}_{\textit{sem}}(\cdot)$ denotes the semantic decoder and $\hat{y}$ denotes the transmitted bitstream (entropy-coded latent representation). 
Additionally, completing the compression loop requires low-level details to support inter-prediction~\cite{sheng2023lvvc}. Therefore, we propose that pixel-oriented reconstruction and machine-oriented reconstruction are conducted separately, with the former providing pixel-level priors for the latter.  

\begin{figure*}[!t]
\centering
\includegraphics[width=0.99\textwidth]{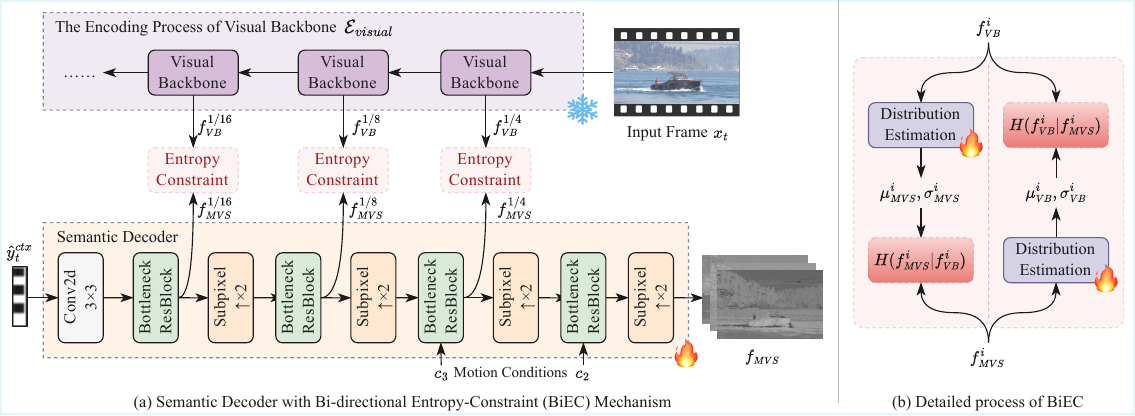}
\caption{ \myrevision{(a) Structure of the proposed information-alignment-based semantic decoder and BiEC mechanism. ``$\uparrow\times 2$'' denotes two times up-sampling. During training, the intermediate features for the decoding $f_{\textit{MVS}}$ are made hierarchically symmetric with the features $f_{\textit{VB}}$ from the visual backbone. (b) Detailed process of BiEC, where $f_{\textit{VB}}^i$ and $f_{\textit{MVS}}^i$ are information-aligned through the bi-directional conditional entropy loss.}}
\label{figure-both-biec}
\end{figure*}

To achieve the above ideas, we propose the Symmetric Entropy-Constrained Video Coding framework for Machine (SEC-VCM), as shown in Fig.~\ref{figure-overview}. Specifically, it consists of three main steps: inter-prediction and contextual coding for the basic compression loop, machine-oriented semantic decoding with the Bi-directional Entropy-Constraint (BiEC) mechanism, and semantic-pixel gated fusion. 
(1)~In the first step, we adopt the basic compression loop from DCVC-HEM~\cite{dcvc_hem}, where motion and contextual information are compressed, quantized, entropy-coded, and then reconstructed. The compensated features $c_1$, $c_2$, and $c_3$ serve as motion priors for both contextual coding and our semantic decoding. The motion latent $\hat{y}_t^{mv}$ and contextual latent $\hat{y}_t^{ctx}$ are then transmitted in the bitstream.
(2)~In the second step, the contextual latent representation $\hat{y}t^{ctx}$ is employed for semantic decoding, where the intermediate feature $f_{\textit{MVS}}$ for MVS-oriented reconstruction is symmetrically aligned with $f_{\textit{VB}}$ extracted by the visual backbone $\mathcal{E}_{\textit{visual}}$. Specifically, in the BiEC mechanism, the symmetry is enforced at three scales, and the alignment is achieved by constraining conditional entropy. 
(3)~In the third step, since $f_{\textit{MVS}}$ mainly contains the semantic information, additional detailed pixel-level priors $f_{\textit{pixel}}$ are injected into the final reconstruction process to enrich frame details, suppress artifacts, and produce frames better suited to MVS. Notably, our framework is general across multiple downstream tasks; therefore, we remove the task loss $\mathcal{L}_{task(\hat{x})}$ that relies on specific labeled datasets. Our paradigm is illustrated as follows:
\begin{equation}
\left\{
\begin{aligned}
    \hat{x}=&\mathcal{D}_{\textit{sem}}(\lfloor\mathcal{E}(x)\rceil|\mathcal{E}_{\textit{visual}}(x),f_{\textit{pixel}}),  \\
    \mathcal{L}=&R + D(x, \hat{x}) + D(\mathcal{E}_{\textit{visual}}(x),\mathcal{E}_{\textit{visual}}(\hat{x}))\\
    &+H(f_{\textit{MVS}}|f_{\textit{VB}})+H(f_{\textit{VB}}|f_{\textit{MVS}}),
\end{aligned}
\right.
\end{equation}
\noindent where $\mathcal{D}_{\textit{sem}}(\cdot)$ denotes the proposed information-alignment-based semantic decoder, $f_{\textit{VB}}$ denotes the feature extracted by the visual backbone, $f_{\textit{MVS}}$ denotes the feature in the semantic decoder, and $f_{\textit{pixel}}$ denotes the feature used for low-level reconstruction. $H(f_{\textit{MVS}}|f_{\textit{VB}})$ and $H(f_{\textit{VB}}|f_{\textit{MVS}})$ denote the conditional entropy terms used as loss items in the proposed BiEC mechanism.

\subsection{Semantic Decoding with Symmetric Entropy-Constraint}

To leverage the visual backbone's strong capability for extracting machine-vision-oriented feature representations, the video decoding process is hierarchically and symmetrically aligned with the encoding process of VB during training. It enables the decoder to inherit the prior knowledge learned by the backbone that has experienced large-scale pretraining, focusing purely on features relevant to machine vision tasks. To achieve this idea, an information-aligned semantic decoder and a BiEC mechanism are designed, as shown in Fig.~\ref{figure-both-biec}.

\subsubsection{Information-Alignment-based Semantic Decoder} The semantic decoder comprises four bottleneck residual blocks and three up-sampling layers. It progressively recovers the quantized and compact latent representation $\hat{y}_{ctx}$ (at $1/16$ resolution) back to the original spatial resolution, producing machine-vision-oriented features $f_{\textit{MVS}}$. The inter-predicted features $c_2$~(at $1/2$ resolution) and $c_3$~(at $1/4$ resolution) are used as conditions for frame reconstruction, referring to~\cite{dcvc, dcvc_hem}. Notably, intermediate features $f_{\textit{MVS}}$ within the semantic decoder are aligned with the intermediate results $f_{\textit{VB}}$ from the visual backbone at three distinct scales. These three hierarchical alignments make the decoder symmetric to the pretrained guidance backbone. 

\subsubsection{Bi-directional Entropy-Constraint Mechanism} 

The features $f_{\textit{VB}}$ extracted by the pretrained backbone are optimized for analysis but not directly suitable for frame reconstruction. Therefore, classic alignment metrics such as MSE or KL-divergence are ineffective when applied between $f_{\textit{VB}}$ and the feature $f_{\textit{MVS}}$ in the video decoding process. 
Instead, we propose to align $f_{\textit{VB}}$ and the reconstruction-oriented feature $f_{\textit{MVS}}$ from an information-theoretic perspective.
The core idea is to ensure that both feature sets carry equivalent information, allowing accurate inference of one from the other. The distribution estimation (DE) module is introduced to estimate the element-wise distribution parameters $(\mu_{\textit{MVS}}, \sigma_{\textit{MVS}})$ of $f_{\textit{MVS}}$ based on $f_{\textit{VB}}$. A more accurate parameter estimation implies a higher conditional probability and thus a lower conditional entropy, indicating that $f_{\textit{VB}}$ retains all information necessary to infer $f_{\textit{MVS}}$, therefore $f_{\textit{MVS}}$ is a subset of $f_{\textit{VB}}$. The above process is formulated as follows:
\begin{equation}
\begin{aligned}
    \mu_{\textit{MVS}}^i, \sigma_{\textit{MVS}}^i &= DE(f_{\textit{VB}}^i),\\
    H(f_{\textit{MVS}}|f_{\textit{VB}})\Downarrow &\Rightarrow \sum_{i}p(f^i_{\textit{MVS}}|\mu^i_{\textit{MVS}},\sigma^i_{\textit{MVS}})\Uparrow,
\end{aligned}
\label{eq1}
\end{equation}
\noindent where $\mu_{\textit{MVS}}^i$ and $\sigma_{\textit{MVS}}^i$ denote estimated element-wise distribution parameters of $f_{\textit{MVS}}^i$. $i \in \{1/4, 1/8, 1/16\}$ denotes the layer-index of $f_{\textit{MVS}}$, $DE(\cdot)$ denotes the distribution estimation module, $H(\cdot|\cdot)$ denotes the conditional entropy, and $p(\cdot|\cdot)$ denotes conditional probability~\cite{balle2016end}. ``$\Rightarrow$'' denotes the transformation of the problem definition. ``$\Downarrow$'' and ``$\Uparrow$'' denote the optimization goal, minimizing the conditional entropy and maximizing the conditional probability, respectively. 
\begin{figure}[!t]
\centering
\includegraphics[width=0.99\linewidth]{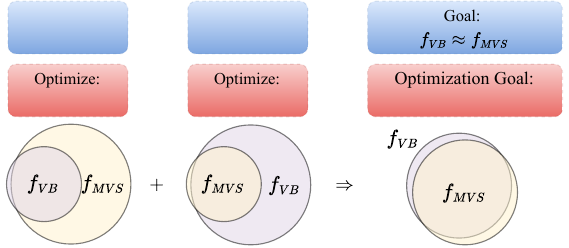}
\caption{The process of the BiEC mechanism. Information alignment between $f_{\textit{MVS}}$ for decoding and $f_{\textit{VB}}$ from the visual backbone is achieved through the constraint of conditional entropy in two distinct directions. ``$\Downarrow$'' indicates that the optimization goal is to minimize the item. }
\label{figure-biec}
\end{figure} 
\begin{figure}[!t]
\centering
\includegraphics[width=0.99\linewidth]{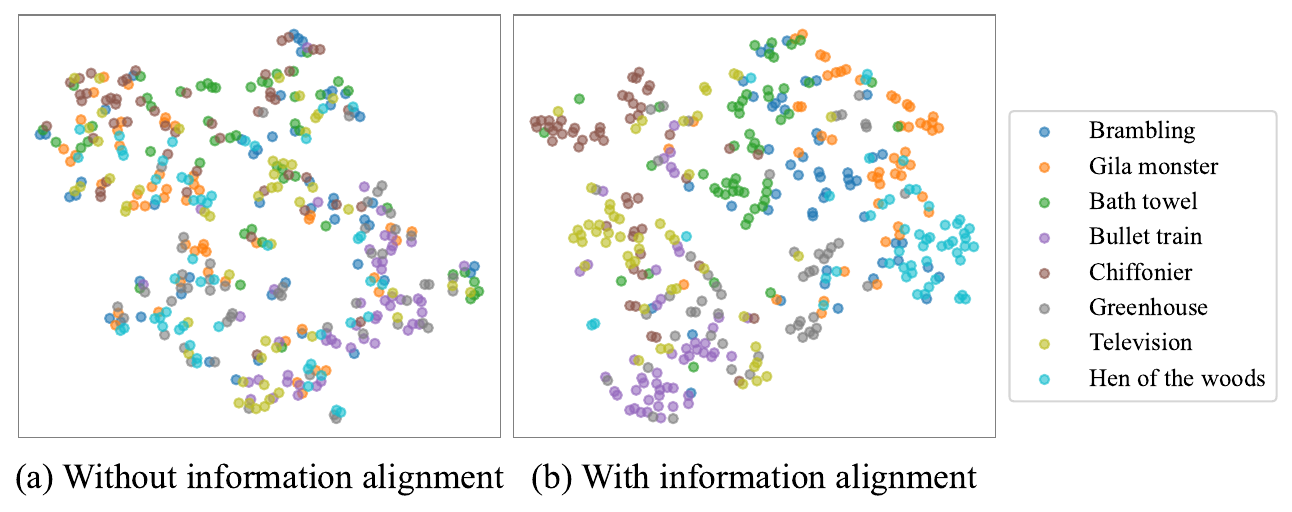}
\caption{The t-SNE visualization of intermediate features $f_{\textit{MVS}}^{1/8}$ in the semantic decoder without and with symmetric information alignment. After alignment via the BiEC mechanism, intra-class distances are effectively reduced, and the intermediate features $f_{\textit{MVS}}^{1/8}$ in the decoding process exhibit more discriminative semantic representations. }
\label{figure-tsne}
\end{figure}

To enforce symmetry, we further impose a conditional entropy constraint in the reverse direction, thus making $f_{\textit{VB}}$ a subset of $f_{\textit{MVS}}$. Through this bi-directional constraint, the features $f_{\textit{MVS}}$ in the semantic decoder and the visual-backbone-extracted representations $f_{\textit{VB}}$ are softly aligned, as illustrated in Fig.~\ref{figure-biec}. The above process is formulated as follows:
\begin{equation}
\begin{aligned}
    \mu_{\textit{VB}}^i, \sigma_{\textit{VB}}^i &= DE(f_{\textit{MVS}}^i),\\
    H(f_{\textit{VB}}|f_{\textit{MVS}})\Downarrow &\Rightarrow \sum_{i}p(f^i_{\textit{VB}}|\mu^i_{\textit{VB}},\sigma^i_{\textit{VB}})\Uparrow ,
\end{aligned}
\label{eq1}
\end{equation}
\noindent where $\mu_{\textit{VB}}^i$ and $\sigma_{\textit{VB}}^i$ denote estimated element-wise distribution parameters of $f_{\textit{VB}}^i$.

Specifically, each DE module contains a convolutional layer and a ResBlock~\cite{resnet50}. Estimated distribution parameters $\mu$ and $\sigma$ have the same shape as the corresponding feature $f$. Following existing methods~\cite{balle2016end, dcvc_hem}, the Laplace distribution is used to model the conditional probability of the features $f_{\textit{MVS}}$ and $f_{\textit{VB}}$. The conditional entropy $H(f_{\textit{VB}}|f_{\textit{MVS}})$ and $H(f_{\textit{MVS}}|f_{\textit{VB}})$ are calculated as follows: 
\begin{equation}
\begin{aligned}
    H(f_{\textit{VB}}|f_{\textit{MVS}})&= \sum_i \sum_j -log_2(p(f_{\textit{VB}}^{i,j}|\mu_{\textit{VB}}^{i,j},\sigma_{\textit{VB}}^{i,j})), \\
    H(f_{\textit{MVS}}|f_{\textit{VB}})&= \sum_i \sum_j -log_2(p(f_{\textit{MVS}}^{i,j}|\mu_{\textit{MVS}}^{i,j},\sigma_{\textit{MVS}}^{i,j})),
\end{aligned}
\label{eq1}
\end{equation}
\noindent where $i$ and $j$ denote the layer and position indices, respectively. 
 
\begin{figure}[!t]
\centering
\includegraphics[width=0.99\linewidth]{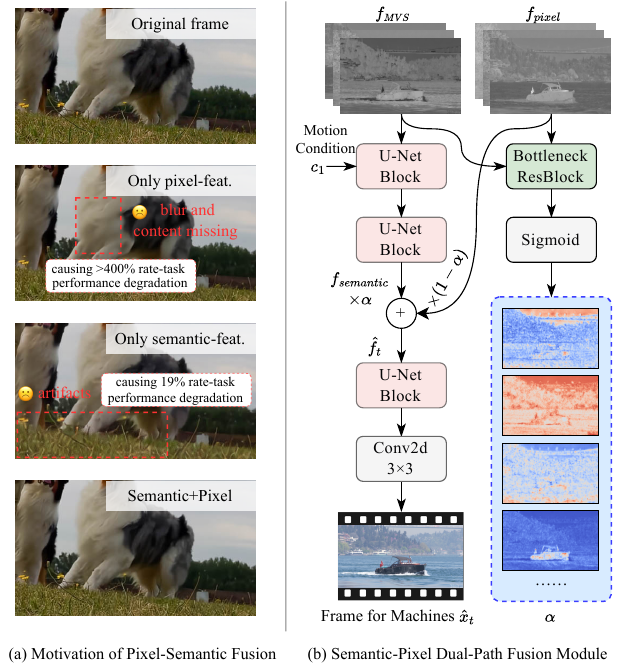}
\caption{\myrevision{(a) Visualization results of reconstructed frame using different feature types at $bpp\approx0.08$. Relying only on pixel-oriented feature $f_{\textit{pixel}}$ tends to miss fine details of the dog's fur texture, while using only semantic-level feature $f_{\textit{semantic}}$ leads to the loss of grass texture and introduces artifacts. By fusing $f_{\textit{semantic}}$ and $f_{\textit{pixel}}$, our method reconstructs the dog's fur details better and avoids artifacts in the grass. (b) Structure of the proposed SPDF module. Machine vision system (MVS)-oriented semantic feature $f_t^{\textit{MVS}}$ and pixel-related feature $f_t^{\textit{pixel}}$ are selectively combined based on the factor $\alpha$.}}
\label{figure-both-spdf}
\end{figure}

\subsubsection{Discussion} Through the BiEC mechanism, the semantic decoder is encouraged to reconstruct semantic representations by relying only on features that are beneficial to video understanding, guided by the visual backbone. In other words, the intermediate features $f_{\textit{MVS}}$ of the semantic decoder are expected to be semantic-aware. To illustrate this property, we provide a visualization. Specifically, 400 images from 8 classes of the ImageNet-1k~\cite{resnet50} validation set are fed into our framework, and the intermediate features $f_{\textit{MVS}}^{1/8}$ of the semantic decoder are projected into a two-dimensional space using t-SNE~\cite{tsne}. Since our framework is designed for P-frames and cannot be directly evaluated on single frames, all pixels of the reference frame are set to zero for this visualization. As shown in Fig.~\ref{figure-tsne}, the semantic decoder, under the joint effect of the visual backbone and the BiEC mechanism, exhibits improved capability in forming compact and semantically meaningful feature representations.

\subsection{Semantic-Pixel Dual-Path Fusion}

The BiEC-based semantic decoder focuses on features relevant to machine vision tasks, while the motion- and context-related modules aim to reconstruct pixel-level details. These two objectives are not fully aligned. Although machine-oriented tasks and traditional distortion metrics (e.g., PSNR, MS-SSIM) both emphasize structural and color preservation, they differ in their sensitivity to low-level textures and patterns.
To bridge this gap, we propose the SPDF module, which injects pixel-level priors into the machine-oriented reconstruction by selectively combining pixel-level features $f_{\textit{pixel}}$ and semantic-level features $f_{\textit{MVS}}$.
Specifically, as illustrated in Fig.~\ref{figure-both-spdf}~(a), using only $f_{\textit{MVS}}$ helps retain the texture of the dog's fur and may increase its detection accuracy, but it ignores grass textures and introduces artifacts, which undermine the semantics of the grass region. In contrast, relying solely on $f_{\textit{pixel}}$ leads to a blur under limited bitrates. By fusing both $f_{\textit{MVS}}$ and $f_{\textit{pixel}}$, our method effectively suppresses grass artifacts and preserves fine textures of fur.

The structure of the SPDF module is shown in Fig.~\ref{figure-both-spdf}(b). Specifically, the machine-oriented feature $f_{\textit{MVS}}$ and pixel-level feature $f_{\textit{pixel}}$ are both fed into the SPDF module. A residual block is employed to compute an element-wise gating coefficient $\alpha$, which is formulated as follows: 
\begin{equation}
\begin{aligned}
    \alpha &= Sigmoid(ResBlock(Concat(f_{\textit{MVS}},f_{\textit{pixel}}))), 
\end{aligned}
\end{equation}
\noindent where $ResBlock(\cdot)$ denotes Bottleneck-ResBlock~\cite{resnet50}, $Concat(\cdot, \cdot)$ denotes channel-wise concatenation. Two stacked U-Nets are employed to perform deeper feature transformations on $f_{\textit{MVS}}$ and get the semantic-level feature $f_{\textit{semantic}}$. Subsequently, $f_{\textit{semantic}}$ and $f_{\textit{pixel}}$ are linearly combined based on the coefficient~$\alpha$. After fusion, another U-Net is used to refine the combined features, and a Conv2D layer is applied to reduce the channel dimension to finally generate the machine-vision-friendly frame $\hat{x}_t$. The above process is formulated as follows: 
\begin{equation}
\begin{aligned}
    & f_{\textit{semantic}} = U(f_{\textit{MVS}},c_1), \\
    & \hat{x}_t = Conv2d(U(\alpha \cdot f_{\textit{semantic}} + (1-\alpha) \cdot f_{\textit{pixel}})),
\end{aligned}
\end{equation}
\noindent where $U(\cdot)$ denotes U-Net(s), $Conv2d(\cdot)$ denotes a 2D-convolutional layer, $\alpha$ denotes the extracted coefficient map for feature fusion, and $c_1$ denotes the motion condition~\cite{dcvc_hem}. 

\subsection{Framework Optimization}

\begin{table}[t]
\renewcommand\arraystretch{1.2} 
\centering
\caption{Training strategy for proposed SEC-VCM framework. ``M'' stands for motion-related modules, ``C'' stands for contextual-coding-related modules, and ``S'' indicates proposed modules for machine-oriented reconstruction. Each epoch contains about 64k iterations. }
\label{table_training_details}
\setlength{\tabcolsep}{5pt}
\begin{tabular}{cccccc}
\toprule
Stage & Modules & $\mathcal{L}$ & $lr$ & GOP & Epochs \\
\midrule
\multirow{7}{*}{Base} & M & $D_{\textit{MSE}}(x,x_{m})$ & 1e-4 & 2 & 1 \\
& C & $D_{\textit{MSE}}(x,\bar{x})$ & 1e-4 & 2 & 1 \\
& C & $D_{\textit{MSE}}(x,\bar{x})$ & 1e-4 & 3 & 1 \\
& C & $D_{\textit{MSE}}(x,\bar{x})$ & 1e-4 & 6 & 1 \\
& M & $D_{\textit{MSE}}(x,x_{m}),R_{m}$ & 1e-4 & 6 & 8 \\
& C & $D_{\textit{MSE}}(x,\bar{x}),R_{c}$ & 1e-4 & 6 & 8 \\
& M,C & $\mathcal{L}_{\textit{base}}$ & 1e-4$\rightarrow$1e-5 & 6 & 20 \\
\midrule
\multirow{5}{*}{VCM} & S & $\mathcal{L}_{e},\mathcal{L}_{\textit{cons}}$ & 1e-4 & 2 & 1 \\
& S & $\mathcal{L}_{e},\mathcal{L}_{\textit{cons}}$ & 1e-4 & 3 & 1 \\
& S & $\mathcal{L}_{e},\mathcal{L}_{\textit{cons}}$ & 1e-4 & 6 & 1 \\
& S & $\mathcal{L}_{e},\mathcal{L}_{\textit{cons}}$ & 1e-4$\rightarrow$1e-5 & 6& 5 \\
& M,C,S & $\mathcal{L}_{\textit{total}}$ & 1e-5 & 6 & 2 \\
\bottomrule
\end{tabular}
\end{table}

Using pixel-domain distortion metrics (e.g., MSE) rather than perception-related metrics is essential in the early stage of codec training~\cite{sheng2023lvvc}. Meanwhile, existing optical flow-based inter-prediction methods are typically optimized for pixel-domain quality instead of MVS. Therefore, we divide the training process into two main stages. As shown in Table~\ref{table_training_details}, a multi-stage optimization strategy is adopted to achieve optimal rate-task performance.

In the base stage, the network is trained to achieve accurate inter-prediction and fine-grained pixel-domain reconstruction. This stage includes five sub-stages that separately optimize the motion-related and contextual-coding-related modules to improve overall rate-distortion (RD) performance, following~\cite{dcvc_tcm, dcvc_hem, sheng2023lvvc}. To enhance perceptual quality, learned perceptual image patch similarity (LPIPS)~\cite{lpips} is also introduced, making the baseline codec more suitable for perception-driven tasks. After this stage, the codec is capable of completing the full compression loop. The rate-distortion loss $\mathcal{L}_{\textit{base}}$ of this stage is formulated as follows:
\begin{equation}
    \mathcal{L}_{\textit{base}} = \lambda_{\textit{MSE}} D_{\textit{MSE}}(x,\bar{x}) + \lambda_{\textit{LPIPS}}D_{\textit{LPIPS}}(x,\bar{x})+R_{m}+R_{c},
\end{equation}
\noindent where $D_{\textit{MSE}}(\cdot,\cdot)$ denotes mean square error (MSE) distortion. $D_{\textit{LPIPS}}(\cdot,\cdot)$ denotes the LPIPS distortion. $R_m$ and $R_c$ denote bits per pixel (bpp) of the motion and context bitstreams, respectively. $\lambda_{\textit{MSE}}$ and $\lambda_{\textit{LPIPS}}$ denote weights for bitrate-distortion balance.  

In the VCM stage, the SEC-VCM framework is optimized for machine vision with the help of visual backbones from two perspectives. First, the VFM is used to achieve hierarchical information symmetry from the entropy perspective (section~III.B), formulated as: 

\begin{equation}
    \mathcal{L}_{e} = H(f_{\textit{VB}}|f_{\textit{MVS}})+H(f_{\textit{MVS}}|f_{\textit{VB}}),
\end{equation}

\noindent where $H(\cdot|\cdot)$ denotes the conditional entropy. Second, multiple types of vision models are employed to maintain consistency between the reconstructed and original videos. Referring to~\cite{free_vsc, smc++, smc}, Spynet~\cite{spynet} is utilized to preserve pixel-level inter-frame relationships in reconstructed video, while DINOv2~\cite{dinov2}, Swin Transformer~\cite{swin_transformer}, and ResNet~\cite{resnet50} are used to maintain machine-vision consistency in the feature domain. The consistency loss $\mathcal{L}_{\textit{cons}}$ is formulated as follows: 
\begin{equation}
    \mathcal{L}_{\textit{cons}} = \sum_i \sum_j \lambda_i\ D_{\textit{MSE}}(\textit{Model}_i^j(x), \textit{Model}_i^j(\hat{x})),
\end{equation}
\noindent where $\textit{Model}_i^j(\cdot)$ denotes the intermediate feature of the $j$-th layer in the $i$-th model, and $\lambda_i$ denotes the balance weight for the $i$-th model. Overall, the whole training strategy can be regarded as a multi-task optimization, as follows: 
\begin{equation}
    \mathcal{L}_{\textit{total}} = \mathcal{L}_{\textit{base}} + \lambda_e\mathcal{L}_e + \lambda_{\textit{cons}}\mathcal{L}_{\textit{cons}},
\end{equation}
\noindent where $\lambda_{e}$ and $\lambda_{\textit{cons}}$ denote the balance weights. 

\section{Experiments}

\subsection{Experimental Settings of Video Codec}

\subsubsection{Settings of Vision Models}

For the BiEC mechanism, Swin Transformer~(base)~\cite{swin_transformer} is employed during the process of training to align information with the proposed semantic decoder. Specifically, the intermediate features from its second, third, and fourth stages are utilized in our framework and named $f_{\textit{VB}}^{1/4}$, $f_{\textit{VB}}^{1/8}$, and $f_{\textit{VB}}^{1/16}$.
To maintain machine-perceptual consistency, $res2$ and $res3$ from both ResNet-18~\cite{resnet50} and Swin Transformer~(tiny)~\cite{swin_transformer} are employed, along with the $4$-th and $8$-th layers of DINOv2~(small)~\cite{dinov2}. Deeper layers in VFM are excluded to prevent overfitting~\cite{free_vsc}. Additionally, the reconstructed frames are twice down-sampled by linear interpolation before being fed into Spynet~\cite{spynet}, to prevent overemphasis on pixel-level information during training. 

\subsubsection{Training Datasets}

The SEC-VCM is trained on the Vimeo-90k dataset~\cite{vimeo90k}. The Vimeo-90k dataset is widely used for training NVC frameworks and contains $89,800$ video sequences with 7 frames. 

\subsubsection{Training Settings}

As shown in TABLE~\ref{table_training_details}, our training process consists of 50 epochs, with 40 epochs in the base stage and 10 epochs in the VCM stage. The training is conducted on four parallel NVIDIA RTX 3090 GPUs with a global batch size of 8, taking about 20 days. Notably, when training with long GOP, we retain gradients across continuous P-frames to better suppress the cumulative errors introduced by the lossy compression loop. Weight $\lambda_{\textit{MSE}}$ is randomly sampled from $[64,512]$. Since $D_{\textit{LPIPS}}$ and $\mathcal{L}_{\textit{cons}}$ are considered distortion losses, their weights $\lambda_{\textit{LPIPS}}$ and $\lambda_{\textit{cons}}$ are set to the same value as $\lambda_{\textit{MSE}}$. The weight of the BiEC mechanism $\lambda_{e}$ is set to $1$. 

\subsection{Experimental Settings of Downstream Tasks}
\label{sec:exp_settings_downstream}

\myrevision{To validate the effectiveness of our SEC-VCM on downstream machine tasks, six types of downstream tasks at different levels of granularity are selected for evaluation, covering both conventional computer vision tasks and MLLM-based tasks.}

\subsubsection{Object Detection}

Faster R-CNN~\cite{faster-rcnn} is implemented through the Detectron2 codebase~\cite{wu2019detectron2} and tested on the YTVIS2019 dataset~\cite{ytvis2019}, which contains 2,883 videos in 40 categories, ranging in resolution from 360p to 1080p. Since Faster R-CNN has no officially released weights on the YTVIS2019 dataset. We fine-tune pretrained Faster R-CNN on the YTVIS2019 dataset using the default configuration\footnote{Configuration of Faster R-CNN-R50: \url{https://github.com/facebookresearch/detectron2/blob/main/configs/COCO-Detection/faster_rcnn_R_50_FPN_1x.yaml}} for the following experiments. 

\subsubsection{Video Instance Segmentation} 

The classical CNN-based model CrossVIS\footnote{Configuration of CrossVIS-R50: \url{https://github.com/hustvl/CrossVIS/blob/main/configs/CrossVIS/R_50_1x.yaml}}~\cite{crossvis} and the Swin Transformer-based model Mask2Former-video\footnote{Configuration of Mask2Former-Video-Swin-Base: \url{https://github.com/facebookresearch/Mask2Former/blob/main/configs/youtubevis_2019/swin/video_maskformer2_swin_base_IN21k_384_bs16_8ep.yaml}}~\cite{cheng2021mask2former} are used to evaluate VIS performance. Notably, we directly adopt their officially released configurations and pretrained models. Experiments are conducted on the YTVIS2019 dataset. 

\subsubsection{Video Object Segmentation} 

We employ XMem~\cite{cheng2022xmem} to evaluate the VOS performance on compressed videos. The DAVIS2017 dataset~\cite{davis2017} is used in our experiments. It contains 150 videos with frame-level VOS annotations. The officially released 1080p videos are used for compression and subsequently resized to 480p for VOS.
 
\subsubsection{Multiple Object Tracking} 

We employ officially released ByteTrack on the MOT17 dataset~\cite{zhang2022bytetrack}, which contains 42 long videos with frame-level annotations and resolutions ranging from 480p to 1080p. 

\subsubsection{\myrevision{Video Grounding}}

\myrevision{We employ Qwen2.5-VL~\cite{bai2025qwen25vltechnicalreport} as the downstream MLLM, and conduct experiments on the YouTube-VIS-2019 validation set~\cite{ytvis2019}. For each annotated instance, its category name serves as the textual query and its tight bounding box serves as the ground truth. All frames of all validation videos are processed, and the average precision (AP) of the predicted bounding boxes is reported.}

\subsubsection{\myrevision{Visual Question Answering}}

\myrevision{We use the same Qwen2.5-VL MLLMs and adopt Video-MME~\cite{videomme} as the benchmark. Due to the prohibitive encoding/decoding cost of traversing the entire benchmark with every compared codec (especially VTM, HM, and Diff-ICMH), we sample 100 short videos and the associated 300 question--answer pairs for evaluation (the sampling list will be released with our code). Evaluation is conducted through VLMEvalKit~\cite{duan2024vlmevalkit}; input videos are sampled to 64 frames, and the overall score is used as the task performance metric.}

\subsection{Comparison Methods}

The proposed SEC-VCM is compared with four types of codecs. (1)~Traditional hybrid codecs, including VTM-23.1~\cite{vtm} and HM-18.0~\cite{hm}. Referring to VCM-related works~\cite{smc++, free_vsc}, these codecs are employed in the Low-Delay-P (LDP) mode. (2)~Open-source neural video coding (NVC) frameworks, such as DCVC-RT~\cite{dcvc_rt}, DCVC-FM~\cite{dcvc_fm}, DCVC-DC~\cite{dcvc_dc}, and DCVC-HEM~\cite{dcvc_hem}. (3)~Human-perception-oriented and generative codecs I$^2$VC~\cite{i2vc}, PLVC~\cite{plvc}, and Control-GIC~\cite{cgic}. (4)~Machine-oriented codec SA-ICM~\cite{sa_icm}, Diff-ICMH~\cite{feng2025difficmh}, and SMC++~\cite{smc++}. The intra-period of PLVC is set to 9 following its official configuration, and the intra-period of other video codecs is set to 32 for fair comparison~\cite{dcvc_hem, dcvc_dc}. All compared methods are implemented through their officially released weights. 

\subsection{Evaluation Metrics}

BD-Rate~\cite{bd_rate} is used to evaluate the rate-task performance of the image and video codecs. It measures the average bitrate difference required to achieve the same task performance relative to an anchor codec, where a lower value indicates better compression efficiency. VTM-23.1 serves as the anchor for calculating BD-Rate. In addition, bits per pixel (bpp) reflects the bitrate consumption of videos. A lower bpp value means a higher compression ratio. Task performance is evaluated by the metrics of each downstream task. 

\begin{table*}[t]   
\renewcommand\arraystretch{1.2}     
\centering   
\caption{\myrevision{BD-rate (\%) $\downarrow$ of compared video coding methods on five downstream machine vision task models. VTM-23.1 is the anchor for calculating BD-rate (\%). All methods are split into four types: (1) standard codecs, (2) neural codecs for distortion in the pixel domain, (3) neural codecs for perception, and (4) neural codecs for machines. The best performance is highlighted with \textbf{bold}, and the runner-up performance is marked with \underline{underline}.}}
\label{table_bd_rate}   
\setlength{\tabcolsep}{2.2pt}   
\begin{tabular}{l|c|ccc|ccc|ccc|ccc|ccc}   
    \toprule   
    \multirow{2}{*}{Methods} & \multirow{2}{*}{Types} & \multicolumn{3}{c|}{Faster R-CNN~\cite{faster-rcnn}} & \multicolumn{3}{c|}{CrossVIS~\cite{crossvis}} & \multicolumn{3}{c|}{Mask2Former~\cite{cheng2021mask2former}} & \multicolumn{3}{c|}{ByteTrack~\cite{zhang2022bytetrack}} & \multicolumn{3}{c}{XMem~\cite{cheng2022xmem}}\\   
    & & AP & AP50 & AP75 & AP & AP50 & AP75 & AP & AP50 & AP75 & MOTA & IDF1 & IDs\textcolor{red}{\textsuperscript{4}} & J\&F & J & F \\
    \midrule   
    VTM-23.1~\cite{vtm} & \multirow{2}{*}{Standard} & 0.0 & 0.0 & 0.0 & 0.0 & 0.0 & 0.0 & 0.0 & \underline{0.0} & 0.0 & 0.0 & 0.0 & 0.0 & 0.0 & 0.0 & 0.0 \\   
    HM-18.0~\cite{hm} & & -2.2 & 1.3 & -1.9 & 7.2 & -15.5 & -8.7 & 19.4 & 82.5 & -21.3 & 11.3 & 32.7 & -85.8 & 21.6 & 28.3 & 15.2 \\   
    \midrule   
    DCVC-HEM (ACM-MM'22)~\cite{dcvc_hem} & & 101.0 & 12.3 & 83.2 & 129.1 & 49.5 & 207.5 & 90.2 & 133.8 & -3.9 & -17.6 & -29.7 & -65.1 & -2.6 & -5.5 & 3.1 \\
    DCVC-DC (CVPR'23)~\cite{dcvc_dc} & NVC & 115.9 & 152.3 & 98.1 & 177.5 & 153.6 & 170.9 & 5.7 & 185.7 & -15.9 &  5.7 & \underline{-44.8} & -78.4 & 4.6 & -12.9 & -6.0 \\
    DCVC-FM (CVPR'24)~\cite{dcvc_fm} & (pixel) & 108.2 & 137.2 & 90.9 & 127.1 & 25.2 & 186.8 & 33.4 & 110.9 & \textbf{-24.8} & \underline{-23.5} & \textbf{-46.8} & \underline{-95.5} & 44.3 & 48.6 & 39.4 \\
    DCVC-RT (CVPR'25)~\cite{dcvc_rt} & & 128.0 & 149.5 & 113.5 & 144.0 & 54.6 & 165.4 & 28.9 & 102.4 & 0.2 & -23.0 & -41.0 & -94.5 & 87.1 & 89.5 & 85.8 \\     
    \midrule   
    PLVC (IJCAI'22)~\cite{plvc} & NVC & 130.9 & 143.3 & 117.0 & 63.4 & 20.9 & \underline{-45.0} & 276.1 & 126.4 & 217.7 & 55.3 & 76.4 & -72.0 & 362.5 & 402.7 & 277.3 \\
    I$^2$VC (ArXiv'24)~\cite{i2vc} & (perception) & 129.9 & 170.3 & 132.3 & 100.9 & 89.9 & 138.5 & 407.7 & 246.8 & 146.8 & 599.1 & 2292.9 & -41.3 & 103.5 & 118.6 & 73.3 \\   
    \midrule   
    SA-ICM (ICIP'24)~\cite{sa_icm} & \multirow{4}{*}{VCM} & 307.1 & 318.6 & 293.9 & 424.3 & 342.9 & 387.1 & 311.3 & 377.4 & 310.7 & 414.4 & 828.7 & \textbf{-182.3} & 160.4 & 165.8 & 156.7 \\
    \myrevision{Diff-ICMH (NeurIPS'25)~\cite{feng2025difficmh}} & & \textbf{\myrevision{-58.8}} & \textbf{\myrevision{-64.6}} & \textbf{\myrevision{-66.0}} & \underline{\myrevision{-20.9}} & \myrevision{-26.2} & \myrevision{-19.1} & \myrevision{95.7} & \myrevision{61.4} & \myrevision{-15.2} & \myrevision{232.2} & \myrevision{79.7} & \myrevision{151.7} & \myrevision{408.5} & \myrevision{425.1} & \myrevision{321.6} \\
    SMC++ (TPAMI'25)~\cite{smc++} & & -13.1 & -4.0 & -20.6 & -18.1 & \underline{-39.0} & -16.0 & \textbf{-32.2} & \textbf{-3.5} & -0.1 & 91.8 & -5.5 & -72.4 & \underline{-18.7} & \textbf{-21.7} & \underline{-16.8} \\
    SEC-VCM (Ours) & & \underline{-46.2} & \underline{-44.4} & \underline{-47.6} & \textbf{-51.2} & \textbf{-64.8} & \textbf{-55.1} & \underline{-23.6} & 62.9 & \underline{-23.0} & \textbf{-44.9} & 17.9 & -86.0 & \textbf{-29.8} & \underline{-19.2} & \textbf{-38.4} \\
    \bottomrule   
\end{tabular}   
\begin{flushleft}
\scriptsize
\textsuperscript{4}\,Experiments show that there is a weak correlation between the IDs metric and bitrate, as detailed in Fig.~\ref{figure-curve-mot}. Therefore, the IDs-based BD-rate does not well reflect the rate-task performance.\\
\end{flushleft}
\end{table*}

\begin{figure*}[!t]
\centering
\includegraphics[width=\textwidth]{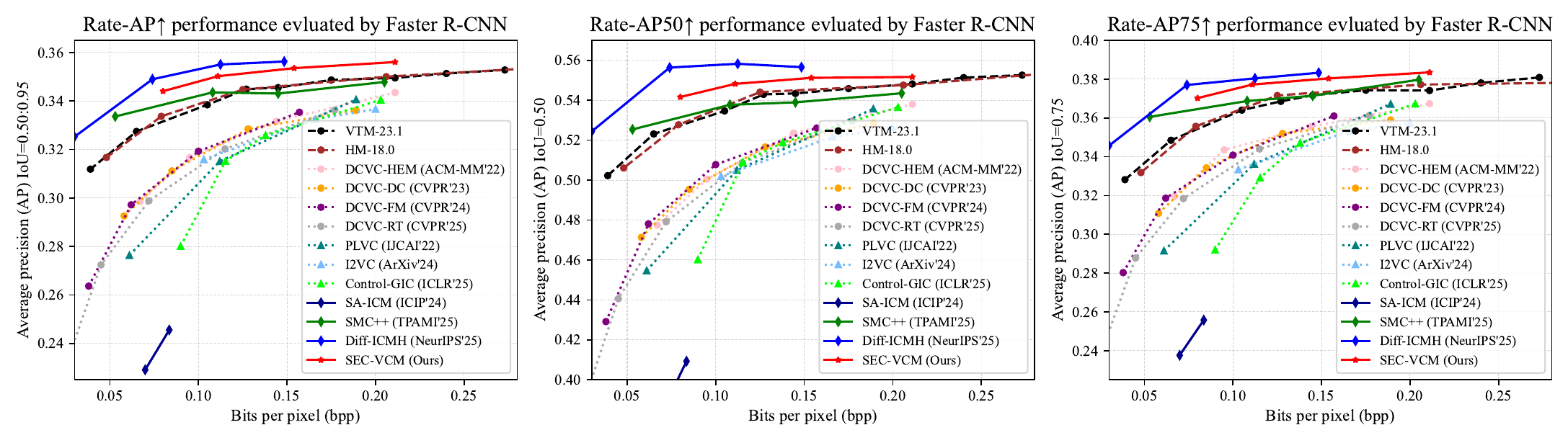}
\vspace{-15pt}
\caption{\myrevision{Rate-task curve of object detection by Faster R-CNN~\cite{faster-rcnn} in terms of AP~$\uparrow$~(1st column), AP50~$\uparrow$~(2nd column), and AP75~$\uparrow$~(3rd column). SEC-VCM achieves the runner-up rate-task performance in all three metrics. }}
\label{figure-curve-od}
\end{figure*}

\begin{figure*}[!t]
\centering
\includegraphics[width=\textwidth]{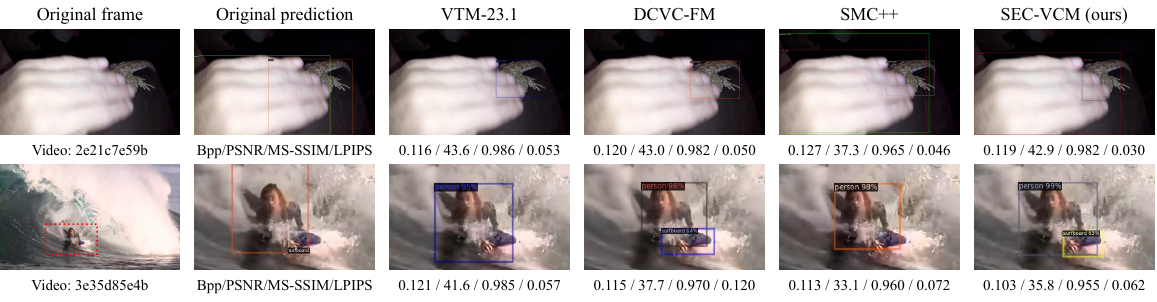}
\vspace{-15pt}
\caption{Visualization of detection results in YTVIS2019 dataset. Faster R-CNN~\cite{faster-rcnn} successfully detects the motion-blurred hand (upper row) and the small surfboard (lower row) from frames that are reconstructed by the proposed SEC-VCM.}
\label{figure-visualization-od}
\end{figure*}

\begin{figure*}[!t]
\centering
\includegraphics[width=\textwidth]{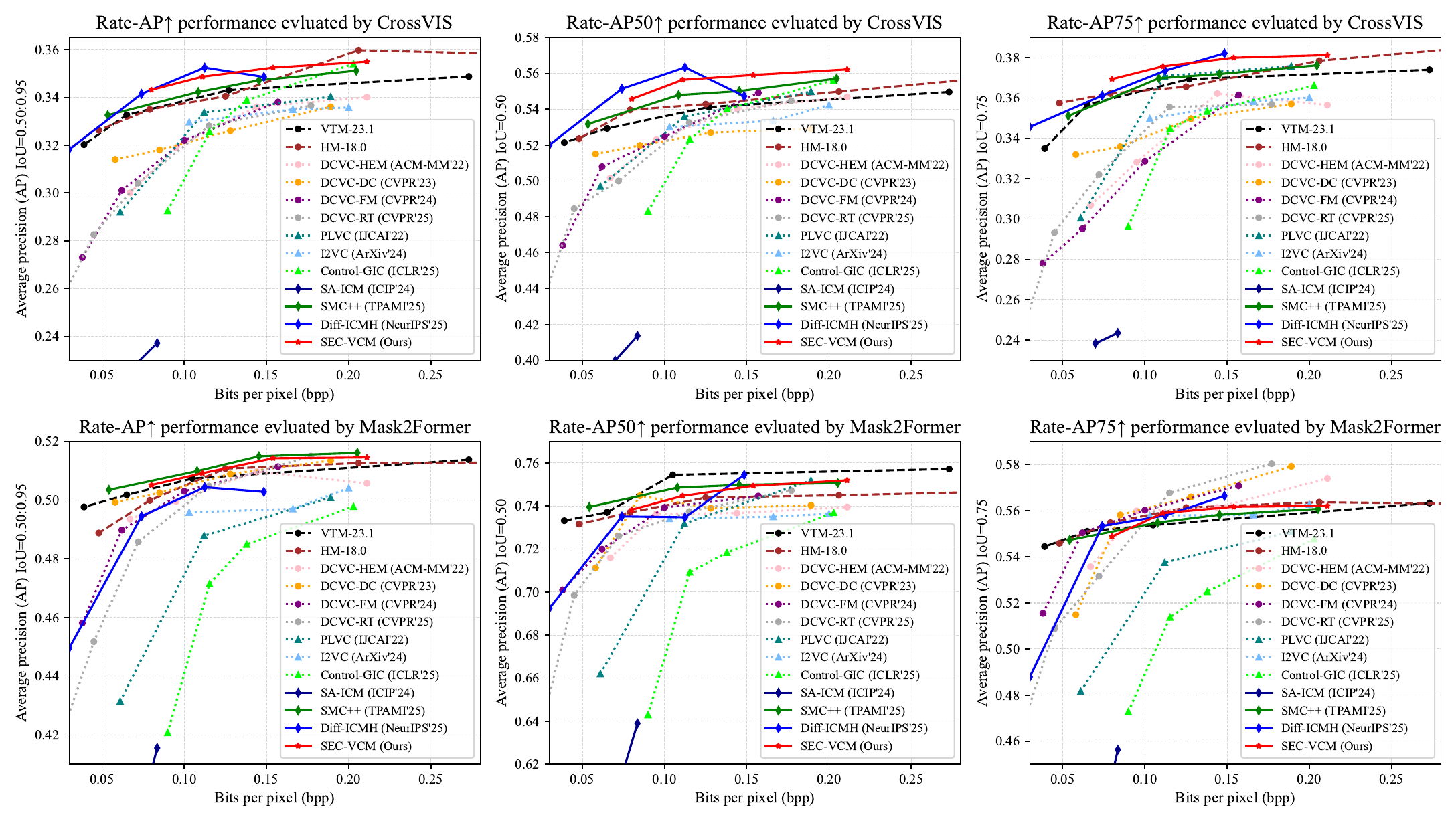}
\vspace{-15pt}
\caption{\myrevision{Rate-task curve of video instance segmentation by CrossVIS~\cite{crossvis} (1st row) and Mask2Former~\cite{cheng2021mask2former} (2nd row) in terms of AP~$\uparrow$~(1st column), AP50~$\uparrow$~(2nd column), and AP75~$\uparrow$~(3rd column). SEC-VCM achieves the best rate-task performance on CrossVIS and secures the runner-up position in rate-AP and rate-AP75 performance on Mask2Former.} }
\label{figure-curve-vis}
\end{figure*}
\subsection{Rate-task Performance and Visualization}

\subsubsection{Object Detection}

The average precision (AP) under three thresholds~\cite{faster-rcnn} is used to evaluate the performance of object detection. \myrevision{As shown in TABLE \ref{table_bd_rate}, the proposed SEC-VCM achieves competitive rate-task performance, ranking second among VCM methods with bitrate reductions of $46.2\%$~(AP), $44.4\%$~(AP50), and $47.6\%$~(AP75) compared to VTM. While Diff-ICMH~\cite{feng2025difficmh} achieves the best performance on this task, SEC-VCM still demonstrates significant improvements over traditional codecs and other neural codecs. The detection performance across various bitrates is shown in Fig.~\ref{figure-curve-od}, where SEC-VCM consistently outperforms most methods except Diff-ICMH.}

We visualize the object detection results using Faster R-CNN~\cite{faster-rcnn}, as shown in Fig.~\ref{figure-visualization-od}. On the reconstructed frames from SEC-VCM, Faster R-CNN successfully detects a motion-blurred hand and a small surfboard, which are missed in the original images and by several competing methods. This improvement is attributed to the fact that the perceptual-oriented reconstruction effectively removes artifacts and distortions that hinder detection.

\begin{figure*}[t]
\centering
\includegraphics[width=\textwidth]{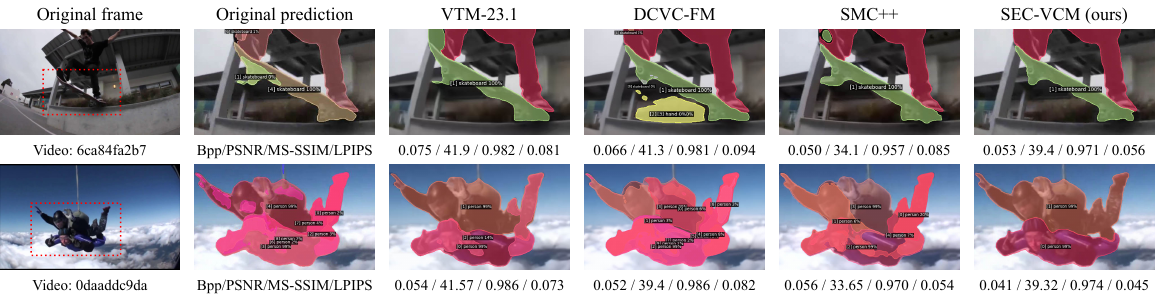}
\vspace{-15pt}
\caption{Visualization on the YTVIS2019 dataset of video instance segmentation results. Under intense motion (1st row) and instance overlap (2nd row), Mask2Former~\cite{cheng2021mask2former} achieves more accurate instance contours on the reconstructed videos from SEC-VCM. }
\label{figure-visualization-vis}
\end{figure*}

\subsubsection{Video Instance Segmentation}

\begin{figure*}[!t]
\centering
\includegraphics[width=\textwidth]{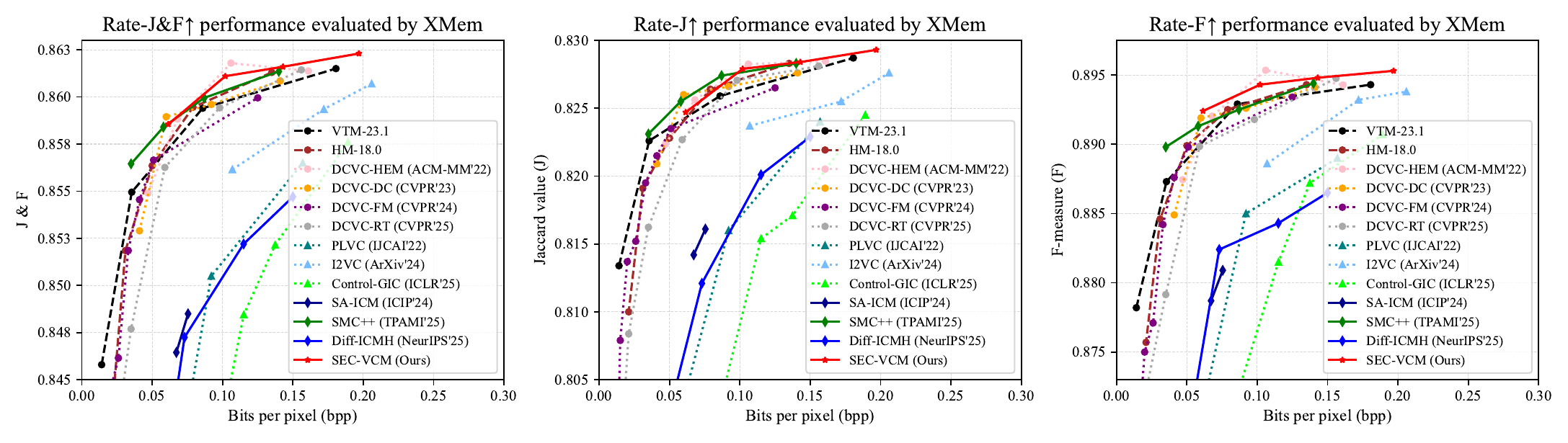}
\vspace{-15pt}
\caption{\myrevision{Rate-task curve of video object segmentation by XMem~\cite{cheng2022xmem}. Task performance is evaluated in terms of J\&F~$\uparrow$~(1st column), J~$\uparrow$~(2nd column), and F~$\uparrow$~(3rd column). SEC-VCM has the best rate-J\&F and rate-F performance and runner-up rate-J performance. }}
\label{figure-curve-vos}
\end{figure*}

\begin{figure*}[!t]
\centering
\includegraphics[width=\textwidth]{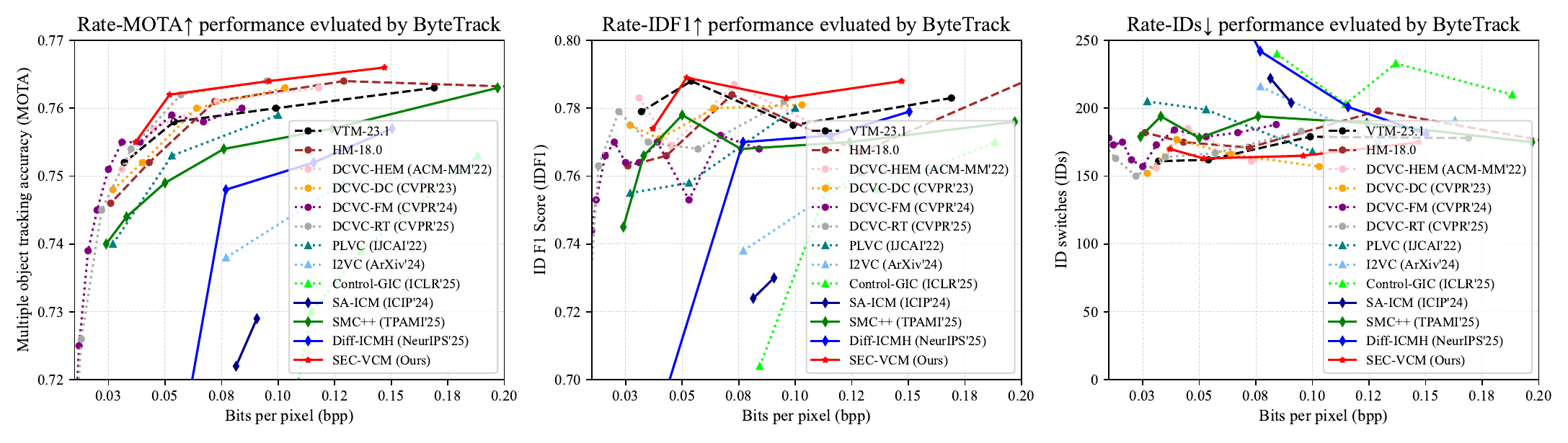}
\vspace{-15pt}
\caption{\myrevision{Rate-task curve of multiple object tracking by ByteTrack~\cite{zhang2022bytetrack}. Task performance is evaluated in terms of MOTA~$\uparrow$~(1st column), IDF1~$\uparrow$~(2nd column), and IDs~$\downarrow$~(3rd column). SEC-VCM achieves the best rate-MOTA performance. }}
\label{figure-curve-mot}
\end{figure*}

AP, AP50, and AP75 are evaluation metrics for VIS~\cite{crossvis, cheng2021mask2former}. As demonstrated in TABLE~\ref{table_bd_rate} and Fig.~\ref{figure-curve-vis}, when CrossVIS is used for instance segmentation, our proposed SEC-VCM achieves the highest compression ratio, saving $51.2\%$~(AP), $64.8\%$~(AP50), and $55.1\%$~(AP75) bitrate compared to VTM. When Mask2Former-video is used, SEC-VCM achieves the 2nd-best results in terms of AP and AP75, while SMC++~\cite{smc++} achieves the best rate-task performance in terms of AP and AP50. 

The visualizations of VIS by Mask2Former~\cite{cheng2021mask2former} are shown in Fig.~\ref{figure-visualization-vis}. For the skateboard video with fast motion in the first row, Mask2Former preserves the object contours more accurately and suppresses incorrect segmentation on the video reconstructed by the proposed SEC-VCM. For the overlapping persons in the second row, it also provides more precise delineation on the reconstructed videos of SEC-VCM.

\subsubsection{Multiple Object Tracking}

Multiple object tracking accuracy~(MOTA), ID F1 score~(IDF1), and ID switches~(IDs) are used to evaluate the performance of tracking, referring to~\cite{zhang2022bytetrack}. As shown in TABLE~\ref{table_bd_rate}, our proposed SEC-VCM outperforms other methods in MOTA and achieves $44.9\%$ bitrate saving compared to VTM. Since IDF1 and IDs are mainly influenced by the inter-frame consistency rather than the frame quality, they show weaker correlation with bitrate. Although SEC-VCM does not have the best rate-task performance in terms of IDF1 and IDs, it outperforms other methods on variable bitrates, as shown in Fig.~\ref{figure-curve-mot}. 

\subsubsection{Video Object Segmentation}

Three metrics---Jaccard value~(J), F-measure~(F), and J\&F---are used to evaluate the performance of video object segmentation. As demonstrated in TABLE~\ref{table_bd_rate} and Fig.~\ref{figure-curve-vos}, our proposed SEC-VCM achieves the most bitrate saving compared to VTM in terms of J\&F~($29.8\%$) and F~($38.4\%$) while achieving a comparable compression ratio with SMC++ on the J metric.  

\subsubsection{Discussion} The above experiments show that the SEC-VCM achieves promising rate-task performance in most metrics. It is worth mentioning that although many methods exhibit excellent rate-distortion (PSNR, MS-SSIM) and rate-perception (LPIPS, etc.) performance that far exceeds standard codecs, their performance in downstream machine vision tasks is not satisfactory. These widely used distortion metrics cannot well reflect the needs of machine vision tasks, which further illustrates the necessity of VCM-related works. Also, developing general and differentiable metrics tailored for downstream tasks remains an urgent and open problem~\cite{iqa_machine_cvpr}.

\begin{table*}[t]
\renewcommand\arraystretch{1.2} 
\centering
\caption{\myrevision{BD-rate (\%) $\downarrow$ of compared VCM methods on two MLLM-based downstream machine vision task models. VTM-23.1 is the anchor for calculating BD-rate (\%). The best performance is highlighted with \textbf{bold}, and the runner-up performance is marked with \underline{underline}.}}
\label{table_bd_rate_mllm}
\setlength{\tabcolsep}{2.2pt}
\begin{tabular}{l|cc|cc}
    \toprule
    \myrevision{Methods} & \multicolumn{2}{c|}{\myrevision{Average precision of video grounding}} & \multicolumn{2}{c}{\myrevision{Score of visual question answering}} \\
    & \myrevision{Qwen2.5-VL-3B-Instruct} & \myrevision{Qwen2.5-VL-7B-Instruct} & \myrevision{Qwen2.5-VL-3B-Instruct} & \myrevision{Qwen2.5-VL-7B-Instruct} \\
    \midrule
    \myrevision{VTM-23.1~\cite{vtm}}& \myrevision{0.0} & \myrevision{0.0} & \myrevision{0.0} & \myrevision{0.0}\\ 
    \myrevision{HM-18.0~\cite{hm}} & \myrevision{-67.6} & \myrevision{-38.1} & \textbf{\myrevision{-45.4}} & \underline{\myrevision{-30.1}} \\
    \midrule
    \myrevision{SA-ICM (ICIP'24)~\cite{sa_icm}} & \myrevision{5.90} & \myrevision{223.3} & \myrevision{48.0} & \myrevision{NaN} \\ 
    \myrevision{Diff-ICMH (NeurIPS'25)~\cite{feng2025difficmh}} & \underline{\myrevision{-83.9}} & \myrevision{59.4} & \myrevision{266.2} & \textbf{\myrevision{-46.9}}\\
    \myrevision{SMC++ (TPAMI'25)~\cite{smc++}} & \myrevision{-83.0} & \underline{\myrevision{-68.7}} & \myrevision{-8.5} & \myrevision{18.1}\\
    \myrevision{SEC-VCM (Ours)} & \textbf{\myrevision{-99.9}} & \textbf{\myrevision{-95.3}} & \underline{\myrevision{-22.4}} & \myrevision{125.6} \\
    \bottomrule
\end{tabular}
\end{table*}

\begin{figure*}[!t]
\centering
\includegraphics[width=\textwidth]{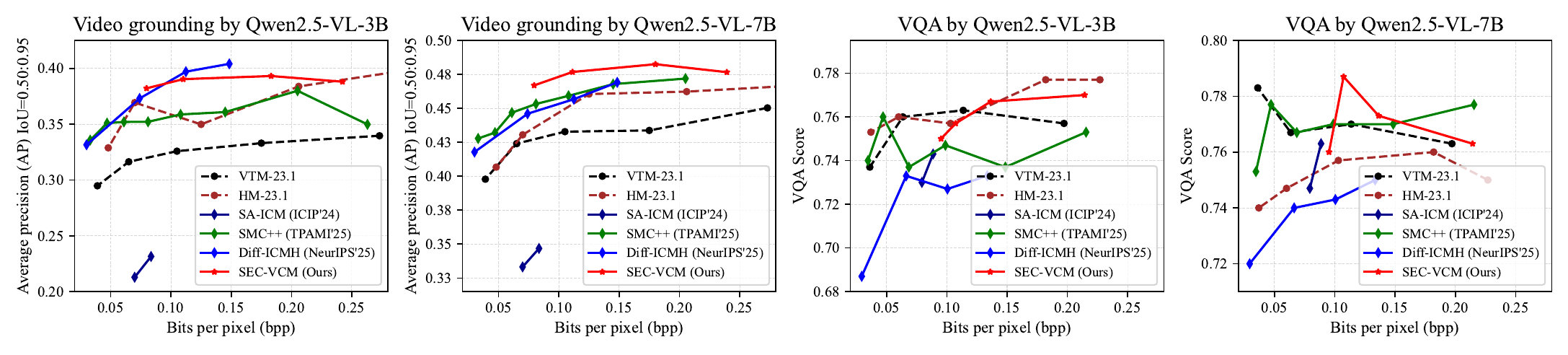}
\vspace{-15pt}
\caption{\myrevision{Rate-task performance on MLLM-based downstream tasks. Left two subplots: video grounding by Qwen2.5-VL. Right two subplots: visual question answering by Qwen2.5-VL. 
}}
\label{figure-curve-mllm}
\end{figure*}

\subsection{\myrevision{Rate-task Performance on MLLM-based Downstream Tasks}}

\subsubsection{\myrevision{Video Grounding}}

\myrevision{The experimental settings of video grounding have been detailed in Section~\ref{sec:exp_settings_downstream}. As shown in TABLE~\ref{table_bd_rate_mllm}, SEC-VCM achieves the best rate-task performance on this task, with remarkable bitrate savings of 99.9\% and 95.3\% over VTM-23.1 on Qwen2.5-VL-3B-Instruct and Qwen2.5-VL-7B-Instruct, respectively. The rate-task curves in Fig.~\ref{figure-curve-mllm} (left) further confirm that SEC-VCM consistently dominates all compared VCM and traditional codecs across the tested bitrate range, for both the 3B and 7B models. These results indicate that the proposed symmetric entropy-constrained decoding effectively preserves the fine-grained spatial cues that are critical for MLLM-based localization, even under aggressive compression.}

\subsubsection{\myrevision{Visual Question Answering}}

\myrevision{The experimental settings of VQA have been detailed in Section~\ref{sec:exp_settings_downstream}. On this task, we observe an interesting phenomenon: the MLLM's score does not necessarily improve monotonically as the bitrate increases, since the strong language prior of MLLMs can largely compensate for compression artifacts so long as the high-level semantics are preserved. Consequently, the rate--task curves of all compared codecs tend to converge to comparable scores, and they are no longer strictly concave. This makes the standard BD-rate metric less representative and less reliable on this task, especially for Qwen2.5-VL-7B-Instruct, where the above phenomenon is more pronounced. We therefore additionally compare the scores at individual bitrate points in Fig.~\ref{figure-curve-mllm} (right). For Qwen2.5-VL-3B-Instruct, SEC-VCM consistently surpasses the VCM baselines SA-ICM and Diff-ICMH across all tested bitrates; for Qwen2.5-VL-7B-Instruct, SEC-VCM outperforms both HM and Diff-ICMH. These observations confirm that SEC-VCM remains competitive on MLLM-based high-level understanding tasks.}

\subsection{Ablation Study}

\begin{table}[!t]
\renewcommand\arraystretch{1.2} 
\centering
\caption{Influence of different backbones on BiEC mechanism. VTM-23.1 is the anchor for calculating BD-rate (\%)~$\downarrow$. The best rate-task performance is highlighted with \textbf{bold}. }
\label{table_ablation_vfm}
\setlength{\tabcolsep}{5.5pt}
\begin{tabular}{lc|ccc}
    \toprule
    Downstream tasks & Metric & SwinTransformer & ResNet & VGGNet \\
    \midrule
    Faster R-CNN~\cite{faster-rcnn} & AP & -46.2 & \textbf{-49.0} & -46.3 \\
    CrossVIS~\cite{crossvis} & AP & \textbf{-51.2} & -32.8 & -30.0 \\
    Mask2Former~\cite{cheng2021mask2former} & AP & -23.6 & -18.9 & \textbf{-35.4} \\
    XMem~\cite{cheng2022xmem} & J\&F & \textbf{-29.8} & 3.8 & 0.1 \\ 
    ByteTrack~\cite{zhang2022bytetrack} & MOTA & \textbf{-44.9} & -23.7 & -20.9 \\
    \midrule
    Average BD-Rate & & \textbf{-39.2} & -24.1 & -26.5 \\ 
    \bottomrule
\end{tabular}
\end{table}

\begin{table}[!t]
\renewcommand\arraystretch{1.2} 
\centering
\caption{Influence of the BiEC loss weight $\lambda_e$. $H$ denotes the value of conditional entropy. VTM-23.1 is the anchor for calculating BD-rate (\%)~$\downarrow$. The best rate-task performance is highlighted with \textbf{bold}. }
\label{table_ablation_lambda}
\setlength{\tabcolsep}{2pt}
\begin{tabular}{lc|ccccc}
    \toprule
    \multirow{2}{*}{Downstream tasks} & \multirow{2}{*}{Metric} & $\lambda_e$=$0.1$ & $\lambda_e$=$0.5$ & $\lambda_e$=$1$ & $\lambda_e$=$5$ & $\lambda_e$=$10$ \\
     & & $H$=$1.73$ & $H$=$1.69$ & $H$=$1.59$ & $H$=$1.50$  & $H$=$1.45$ \\
    \midrule
    Faster R-CNN~\cite{faster-rcnn} & AP & -40.2 & -45.8 & \textbf{-46.2} & -41.0 & -44.4 \\
    CrossVIS~\cite{crossvis} & AP & -31.5 & \textbf{-54.4} & -51.2 & -38.8 & -46.5 \\
    Mask2Former~\cite{cheng2021mask2former} & AP & -33.3 & -11.6 & -23.6 & \textbf{-33.9} & -27.2 \\
    XMem~\cite{cheng2022xmem} & J\&F & -16.9 & 1.4 & -19.2 & \textbf{-38.4} & -25.9 \\ 
    ByteTrack~\cite{zhang2022bytetrack} & MOTA & -10.4 & -29.5 & \textbf{-44.9} & -16.5 & -9.7 \\
    \midrule
    Average BD-Rate & & -24.5 & -26.7 & \textbf{-37.0} & -33.7 & -30.8 \\
    \bottomrule
\end{tabular}
\end{table} 

\begin{table}[!t]
\renewcommand\arraystretch{1.2} 
\centering
\caption{Ablation study on the BiEC-related components.}
\label{table_ablation_biec}
\setlength{\tabcolsep}{4.5pt}
\begin{tabular}{c|ccc|c}
    \toprule 
    Models & $H(f_{\textit{MVS}}|f_{\textit{VB}})$ & $H(f_{\textit{VB}}|f_{\textit{MVS}})$ & Multi-layer & Average bitrate \\ 
    \midrule
    Ours & \myright & \myright & \myright & 0\%\\
    \midrule
    $M_1$ & \myright & \mywrong & \myright & +16.0\%\\
    $M_2$ & \mywrong & \myright & \myright & +7.4\%\\
    \midrule
    $M_3$ & \myright & \myright & \mywrong & +11.3\% \\
    $M_4$ & \myright & \mywrong & \mywrong & +26.9\% \\
    $M_5$ & \mywrong & \myright & \mywrong & +21.5\%\\
    \midrule
    $M_6$ & MSE & MSE & \myright & +21.1\%\\ 
    $M_7$ & $\text{KL}_{\text{channel}}$ & $\text{KL}_{\text{channel}}$ & \myright & +17.3\%\\ 
    $M_8$ & $\text{KL}_{\text{spatial}}$ & $\text{KL}_{\text{spatial}}$ & \myright & +13.5\%\\ 
    \bottomrule
\end{tabular}
\end{table}

\begin{table}[!t]
\renewcommand\arraystretch{1.2} 
\centering
\caption{Ablation study on the semantic-pixel dual-path fusion.}
\label{table_ablation_dualfusion}
\setlength{\tabcolsep}{6pt}
\begin{tabular}{c|ccc|c}
    \toprule
    Models & Gated fusion & Concatenation & Addition & Average bitrate \\
    \midrule
    Ours & \myright & \mywrong & \mywrong & 0\% \\
    \midrule
    $M_9$ & \mywrong & \myright & \mywrong & +13.4\% \\
    $M_{10}$ & \mywrong & \mywrong & \myright & +8.9\% \\
    $M_{11}$ & \mywrong & \mywrong & \mywrong & +19.5\% \\
    \bottomrule
\end{tabular}
\end{table}

\begin{table*}[!h]
\renewcommand\arraystretch{1.2} 
\centering
\caption{\myrevision{Complexity of neural codecs in terms of inference time, encoding time, decoding time, \myrevision{theoretical encoding/decoding complexity (kMACs/pixel),} and number of optimized parameters. The times are evaluated on 1080p videos by a single NVIDIA RTX 3090. }}
\label{table_complexity}
\setlength{\tabcolsep}{6pt}
\begin{tabular}{l|ccc|cc|c}
    \toprule
    \multirow{2}{*}{Methods} & Inference & Encoding & Decoding & \myrevision{Encoding} & \myrevision{Decoding} & Optimized \\
     & (ms) & (ms) & (ms) & \myrevision{(kMACs/pixel)\textcolor{red}{\textsuperscript{5}}} & \myrevision{(kMACs/pixel)} & params\\
    \midrule
    DCVC-HEM (ACM-MM'22)~\cite{dcvc_hem} & 566 & 582 & 209 & \myrevision{1,671} & \myrevision{1,247} & 17.5M \\
    DCVC-DC (CVPR'23)~\cite{dcvc_dc} & 670 & 701 & 528 & \myrevision{1,361} & \myrevision{932} & 19.8M \\
    DCVC-FM (CVPR'24)~\cite{dcvc_fm} & 654 & 684 & 537 & \myrevision{1,153} & \myrevision{881} & 18.3M \\
    DCVC-RT (CVPR'25)~\cite{dcvc_rt}\textcolor{red}{\textsuperscript{6}} & 81 & 36 & 39 & \myrevision{142} & \myrevision{167} & 20.7M \\
    \midrule
    PLVC (IJCAI'22)~\cite{plvc}\textcolor{red}{\textsuperscript{7}} & / & / & / & \myrevision{1,219} & \myrevision{730} & 19.2M \\
    I$^2$VC (ArXiv'24)~\cite{i2vc} & 1030 & / & / & \myrevision{/} & \myrevision{/} & 78.4M \\
    Control-GIC (ICLR'25)~\cite{cgic} & / & 847 & 1285 & \myrevision{7,705} & \myrevision{445} & 131.0M\\
    \midrule
    SA-ICM (ICIP'24)~\cite{sa_icm} & 720 & 647 & 768 & \myrevision{832} & \myrevision{1,194} & 76.5M \\
    \myrevision{Diff-ICMH (NeurIPS'25)~\cite{feng2025difficmh}}\textcolor{red}{\textsuperscript{8}}  & \myrevision{/} & \myrevision{932} & \myrevision{59,639} & \myrevision{2,254} & \myrevision{172,858} & \myrevision{75.9M}\\
    SMC++ (TPAMI'25)~\cite{smc++} & 1188 & / & / & \myrevision{/} & \myrevision{/} & 96.2M \\ 
    SEC-VCM (Ours) & 707 & 584 & 411 & \myrevision{1,671} & \myrevision{2,013} & 31.9M \\
    \bottomrule
\end{tabular}
\begin{flushleft}
\scriptsize
\myrevision{\textsuperscript{5}\, We test all methods' kMACs/pixel through fvcore. Since SMC++ is a hybrid structure combining VVenC (an open-source and fast version of H.266/VVC) with neural modules, and I$^2$VC does not decouple its compression and decompression procedures, their theoretical complexities cannot be directly calculated. }\\
\textsuperscript{6}\,We turn off its self-implemented CUDA operations and only enable PyTorch-based operations for fair comparison.\\
\myrevision{\textsuperscript{7}\,PLVC is tested on CPU due to CUDA incompatibility with the RTX~3090, and its kMACs/pixel are GOP-averaged over one HiFiC I-frame and eight RLVC-style P-frames (GOP~$=9$, bi-IPPP). The GAN discriminator is excluded from both the kMACs and the parameter count as it is used only at training time.}\\
\myrevision{\textsuperscript{8}\,The reported kMACs/pixel correspond to 50 diffusion iterations under the default inference setting.}
\end{flushleft}
\end{table*}

\begin{figure}[!t]
\centering
\includegraphics[width=\linewidth]{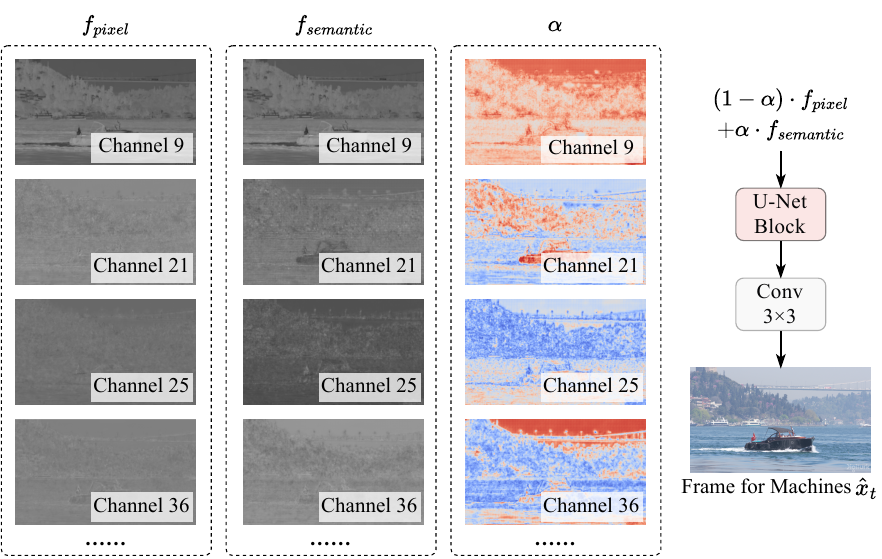}
\caption{Visualization results of the gated fusion process. \textcolor{red}{Red} and \textcolor{blue}{blue} indicate high and low feature values, respectively. The pixel-oriented feature $f_{\textit{pixel}}$ and machine-oriented feature $f_{\textit{semantic}}$ are selectively combined based on $\alpha$ for the final reconstruction. }
\label{figure-spdf}
\end{figure}

\subsubsection{Ablation on the Guidance Visual Backbone in BiEC Mechanism}

While many commonly used backbones have demonstrated strong performance across a wide range of computer vision tasks, we integrate different types of backbones into the proposed BiEC mechanism. Since video reconstruction is a hierarchical up-sampling process, three backbones with hierarchical feature extraction capabilities---Swin Transformer~\cite{swin_transformer}, ResNet~\cite{resnet50}, and VGGNet~\cite{vgg}---are employed in the BiEC mechanism. The evaluation is conducted on the primary metrics of five downstream tasks. As shown in Table~\ref{table_ablation_vfm}, the results reveal notable performance differences across frameworks. The Swin Transformer is eventually chosen as the guidance backbone in the BiEC mechanism, as it significantly outperforms ResNet and VGGNet in three downstream tasks.

\subsubsection{Ablation on the BiEC Mechanism}

Although visual backbones exhibit generalization ability, it is crucial to prevent the entire codec from overfitting to a specific backbone. To evaluate the impact of different levels of the BiEC mechanism on downstream task performance, we train the SEC-VCM with different $\lambda_e$, as shown in Table~\ref{table_ablation_lambda}. Finally, $\lambda_e$ is set to achieve a good trade-off across multiple downstream tasks.

We conduct ablation studies on the bi-directional design and the multi-layer design, as shown in Table~\ref{table_ablation_biec}. Results from $M_1$ and $M_2$ demonstrate that each direction of the entropy constraints plays a critical role in improving rate-task performance of 11.7\%. Also, the experimental results of $M_3$, $M_4$, and $M_5$ show that removing multi-layer design results in significant rate-task performance degradation of 12.0\%. 

Notably, in $M_6$, the proposed conditional entropy loss $\mathcal{L}_e$ is replaced by MSE to perform direct element-wise alignment, resulting in a 21.1\% decrease in rate-task performance. We also applied channel-wise ($M_7$) and spatial-wise ($M_8$) KL divergence losses to achieve distribution-level alignment, which led to performance drops of 17.3\% and 13.5\%, respectively. These results highlight the superiority of the proposed information-level alignment over direct (MSE) and distribution-level (KL divergence) alignment.
\subsubsection{Ablation on the SPDF Module}

To explore how to effectively utilize pixel-oriented features $f_{\textit{pixel}}$ for MVS-oriented reconstruction, channel-wise concatenation ($M_9$) and direct addition ($M_{10}$) are employed to replace the gated fusion, as shown in Table~\ref{table_ablation_dualfusion}. Experimental results demonstrate that gated fusion outperforms the other two fusion methods. Notably, the rate-task performance drops significantly by $19.5\%$ when $f_{\textit{pixel}}$ is not used ($M_{11}$). Furthermore, visualization results indicate that the gated fusion mechanism can selectively combine $f_{\textit{semantic}}$ and $f_{\textit{pixel}}$ for final reconstruction. As shown in Fig.~\ref{figure-spdf}, $f_{\textit{semantic}}$ plays the major role in the reconstruction on channel 9, while contributing only features related to boats, forests, lake surface, and sky on channels 21, 25, and 36.

\subsection{Complexity Analysis}

We compare the inference time, encoding time, decoding time, and the number of optimized parameters of the proposed SEC-VCM with other neural codecs~\cite{dcvc_hem, dcvc_dc, dcvc_fm, i2vc, cgic, smc++, sa_icm}, as shown in Table~\ref{table_complexity}. The inference time reflects the overall computational complexity of the model, excluding file I/O. The encoding and decoding times represent the complexity at the encoder and decoder sides, respectively. The inference, encoding, and decoding times are recorded on 1080p videos and a single NVIDIA RTX 3090 with a batch size of 1. In terms of inference, encoding, and decoding times, our method is comparable to existing high-performance neural video codecs~\cite{dcvc_hem, dcvc_dc, dcvc_fm} and significantly faster than existing human-perception-oriented codecs~\cite{i2vc, cgic} and machine-vision-oriented codecs~\cite{sa_icm, smc++}. In terms of the number of optimized parameters, our framework is lighter than machine-vision-oriented codecs SA-ICM~\cite{sa_icm} and SMC++~\cite{smc++}. Still, it is larger than other neural video codecs due to additional modules (semantic decoder and SPDF module) for machine-oriented reconstruction. 

\myrevision{Notably, our proposed semantic decoder and SPDF module are placed outside the compression-loop, so the whole framework maintains the same encoding complexity as the baseline DCVC-HEM ($1,671$ kMACs/pixel); the additional cost therefore appears only on the decoder side ($2,013$ kMACs/pixel, $+61\%$). Additionally, the pre-trained visual backbone in our framework is only used during the training stage for entropy-constraint and information alignment, adding no burden to the encoding and decoding processes after training is finished.}

\begin{figure*}[!tbp]
    \centering
    \includegraphics[width=\linewidth]{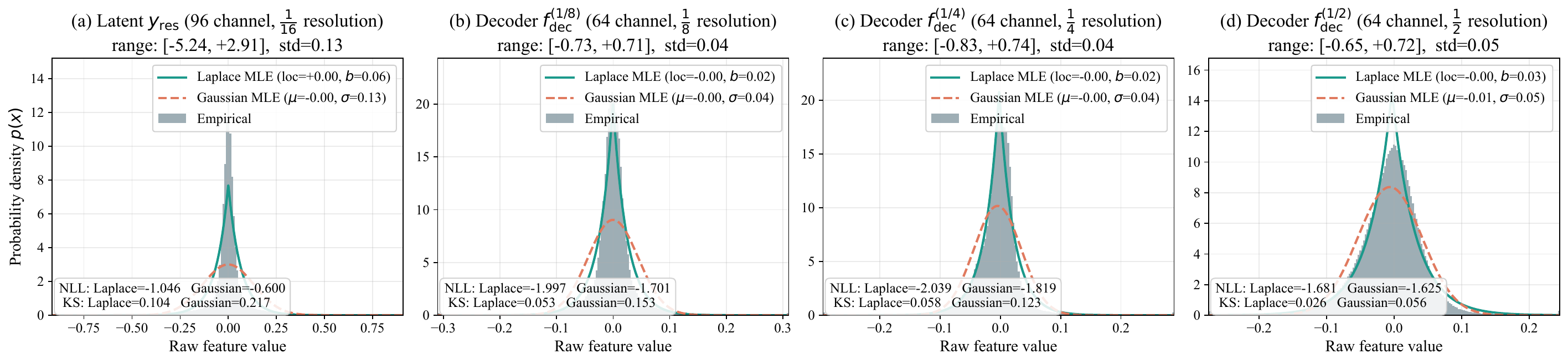}
    \caption{\myrevision{Distribution of the quantized latent $y_{\mathrm{res}}$ and three intermediate decoder features of the unmodified DCVC-HEM baseline, with maximum-likelihood Laplace and Gaussian fits. On NLL and KS, Laplace consistently fits the features better than Gaussian, which justifies its use in our BiEC mechanism.}}
    \label{figure-laplace-vs-gaussian}
\end{figure*}

\subsection{\myrevision{Why Choosing the Laplace Distribution for BiEC}}
\label{subsec_why_laplace}

\myrevision{Transform-domain coefficients of natural images are heavy-tailed and sparse~\cite{wainwright_nips1999}, and a Laplace fits such statistics better than a Gaussian. Also, the Laplace distribution is widely used for entropy coding in recent neural codecs, including the DCVC family~\cite{dcvc, dcvc_tcm, dcvc_hem}, and other DCVC-based works~\cite{dcvc_b, dcvc_sdd}. }

\myrevision{To further verify this choice for our setting, we visualize the empirical distribution of the latent $y_{\mathrm{res}}$ and three intermediate decoder features from the DCVC-HEM baseline without BiEC mechanism, normalization, and regularization in Fig.~\ref{figure-laplace-vs-gaussian}. We compare maximum-likelihood Laplace and Gaussian fits, and use two metrics: the negative log-likelihood (NLL, lower is better, equal to the entropy-coding bit cost up to a constant)~\cite{maximum_likelihood_estimation} and the Kolmogorov--Smirnov statistic (KS, lower is better)~\cite{ks_test_original}. On both metrics and across all four features, Laplace fits the baseline features better than Gaussian, before any BiEC constraint is applied. This confirms that Laplace is the natural choice for our conditional entropy model in BiEC.}

\section{Conclusion}

In this paper, we propose a unified Symmetric Entropy-Constrained Video Coding framework for Machines (SEC-VCM). By aligning the video coding process with the video understanding process of the visual backbone from an entropy perspective, SEC-VCM eliminates machine-vision-irrelevant information. The proposed BiEC mechanism softly maintains consistency between the visual backbone's encoding process and the video decoding process, while the semantic-pixel dual-path fusion module introduces pixel-level priors to suppress artifacts and improve downstream task performance. Extensive experiments on diverse downstream tasks demonstrate the superiority and generalization ability of our method in achieving MVS-friendly reconstruction. 

Nevertheless, our approach still has limitations. Developing a more lightweight framework will be a direction for future work. We believe our proposed method offers insights into bridging the gap between neural video coding and machine vision, serving as a step forward for advancing the field of video coding for machines.

\bibliographystyle{IEEEtran}
\bibliography{references}

\end{document}